\documentclass[11pt,a4paper]{article}
\usepackage{jcappub}
\usepackage{euscript}
\usepackage{amsfonts,latexsym}

\newcommand{\be}[1]{\begin{equation}\label{#1}}
\newcommand{\ee}{\end{equation}}
\newcommand{\ba}[1]{\begin{eqnarray}\label{#1}}
\newcommand{\ea}{\end{eqnarray}}
\newcommand{\rf}[1]{(\ref{#1})}
\newcommand{\nn}{\nonumber}

\begin{document}

\title{Dynamics of astrophysical objects against the cosmological background}

\author[a,b,c]{Maxim Eingorn,} \author[d]{Alexandra Kudinova} \author[a]{and Alexander Zhuk}

\affiliation[a]{Astronomical Observatory, Odessa National University,\\ Dvoryanskaya st. 2, Odessa 65082, Ukraine}

\affiliation[b]{Department of Theoretical and Experimental Nuclear Physics,\\ Odessa National Polytechnic University,\\ Shevchenko av. 1, Odessa 65044, Ukraine}

\affiliation[c]{North Carolina Central University,\\ Fayetteville st. 1801, Durham, North Carolina 27707, USA}

\affiliation[d]{Department of Theoretical Physics, Odessa National University,\\ Dvoryanskaya st. 2, Odessa 65082, Ukraine}

\emailAdd{maxim.eingorn@gmail.com} \emailAdd{autumnforever1@gmail.com} \emailAdd{ai.zhuk2@gmail.com}

\abstract{In this paper, we consider dynamical behavior of astrophysical objects such as galaxies and dwarf galaxies taking into account both the gravitational
attraction between them and the cosmological expansion of the Universe. First, we obtain the general system of equations and apply them to some abstract systems of
galaxies. Then we investigate the collision between the Milky Way and Andromeda in future. Here, we distinguish two models. For the first one, we do not take into
account the influence of the Intra-Group Matter (IGrM). In this case, we demonstrate that for currently known parameters of this system the collision is hardly plausible
because of the angular momentum. These galaxies will approach the minimum distance of about 290 Kpc in 4.44 Gyr from present, and then begin to run away irreversibly
from each other. For the second model, we take into account the dynamical friction due to the IGrM. Here, we find a characteristic value of the IGrM particle velocity
dispersion $\tilde \sigma = 2.306$. For $\tilde \sigma \leq 2.306$, the merger will take place, but for the bigger values of $\tilde\sigma$ the merger can be
problematic. If the temperature of the IGrM particles is $10^5$ K, then this characteristic value of $\tilde\sigma$ corresponds to the IGrM particle mass 17 MeV.
Therefore, for the IGrM particles with masses less than 17 MeV the merger becomes problematic. We also define the region in the vicinity of our Local Group where the
formation of the Hubble flows starts. For such processes, the zero-acceleration surface (where the gravitational attraction is balanced by the cosmological accelerated
expansion) plays the crucial role. We show that such surface is absent for the Local Group. Instead, we find two points and one circle with zero acceleration.
Nevertheless, there is a nearly closed area around the MW and M31 where the absolute value of the acceleration is approximately equal to zero. The Hubble flows are
formed outside of this area.}


\maketitle

\flushbottom


\section{\label{sec:1}Introduction}

\setcounter{equation}{0}

The progress in modern observational cosmology at scales much smaller than the cell of uniformity size (see, e.g., \cite{Sand1,Kar et al,Kar2003,Sand2,Kar2008,Kar2012})
enables to use the new observational data to test different cosmological models. With the help of these data, we can reconstruct the history of galaxies, groups and
clusters of galaxies as well as to predict their future. For example, we can explain the formation of the Hubble flows in the vicinity of the group of galaxies
\cite{Chernin 2012} or predict a possible collision of the Milky Way and Andromeda in future \cite{CoxLoeb}.

According to recent astronomical observations, there is no clear evidence of spatial homogeneity up to sizes  $\sim$ 150 Mpc \cite{Labini}. Deep inside of such scales
and on late stages of evolution, the Universe consists of a set of discrete inhomogeneities (galaxies, groups and clusters of galaxies) which disturb the background
Friedmann Universe\footnote{Of course, such objects as galaxies have their own structure. However, at distances much bigger than the characteristic size of these
objects, we can consider them as discrete inhomogeneities.}. Hence, classical mechanics of discrete objects provides more adequate approach than hydrodynamics with its
continuous flows. In our previous paper \cite{EZcosm1}, we have elaborated this approach for an arbitrary number of randomly distributed inhomogeneities on the
cosmological background and found a gravitational potential of this system. We have shown that this potential has the most natural form in the case of the
Friedmann-Robertson-Walker metrics with the hyperbolic space. Therefore, having the gravitational potential of an arbitrary system of inhomogeneities, we can investigate
their motion taking into account both the gravitational attraction between them and the cosmological expansion of the Universe.

In the present paper, we continue this investigation. First, we obtain the general system of equations of motion for such system and apply these equations to abstract
groups of galaxies to show the effects of gravitational attraction and cosmological expansion. Then, we consider our Local Group to investigate the mutual motion of the
Milky Way and Andromeda. Here, we distinguish two different models. For the first one, we do not take into account the influence of the Intra-Group Matter (IGrM).
Contrary to the conclusions of the paper \cite{CoxLoeb}, we show in this case that for currently known parameters of this system, the collision is hardly plausible in
future because of the angular momentum. These galaxies will approach the minimum distance of about 290 Kpc in 4.44 Gyr from present, and then begin to run away
irreversibly from each other. For the second model, we take into account the dynamical friction due to the IGrM. Here, we find a characteristic value of the IGrM
particle velocity dispersion $\tilde \sigma = 2.306$. For $\tilde \sigma \leq 2.306$, the merger will take place but for bigger values of $\tilde\sigma$ the merger can
be problematic because the galaxies approach a region where the dragging effect of the dynamical friction can be too small to force the galaxies to converge. If the
temperature of the IGrM particles is $10^5$ K, then this characteristic value of $\tilde\sigma$ corresponds to the IGrM particle mass 17 MeV. Therefore, for lighter
masses (and, accordingly, larger values of $\tilde\sigma$) the merger becomes problematic.

Then, we define the region in the vicinity of our Local Group where the formation of the Hubble flows starts. For such
processes, the zero-acceleration surface (where the gravitational attraction is balanced by the cosmological accelerated expansion) plays the crucial role. We take into
account the geometry of the system consisting of two giant galaxies (MW and M31) at the distance 0.78 Mpc. Obviously, if this surface exists, it does not have a
spherical shape. We show that such surface is absent for the Local Group. Instead, we find two points and one circle with zero acceleration. Nevertheless, there is a
nearly closed area around the MW and M31 where the absolute value of the acceleration is approximately equal to zero. The Hubble flows are formed outside of this area.

One of the main conclusions of our work is that cosmological effects become significant already at the scale of the Local Group, i.e. of the order of 1 Mpc.
Therefore, we should take them into account when we consider the dynamics of the Local Group at these distances.

The paper is organized as follows. In section 2 we obtain the general system of equations of motion for arbitrary distributed inhomogeneities in the open Universe. We
apply these equations to abstract systems of galaxies consisting of three and four galaxies in section 3. In section 4, we investigate the mutual motion of the Milky Way
and Andromeda in future. In section 5, we define the zero-acceleration region for our Local Group. The main results are summarized in concluding section 6.


\section{\label{sec:2}General setup}

\setcounter{equation}{0}

In our recent paper \cite{EZcosm1}, we have shown that the "comoving" gravitational potential for a system of gravitating masses $m_i$ is
\be{2.1}
\varphi=-G_N\sum\limits_i m_{i}\frac{\exp(-2l_i)}{\sinh l_i}+\frac{4\pi G_N\overline\rho}{3}\, ,
\ee
where $G_N$ is the Newtonian gravitational constant, $\overline \rho=\mbox{const}$ is the comoving average rest mass density and $l_i$ denotes the comoving geodesic
distance between the i-th mass $m_{i}$ and the point of observation in the open Universe, i.e. in the hyperbolic space. This formula has a number of advantages with
respect to the flat and spherical space cases \cite{EZcosm1}. First, this potential is finite at any point of space (excluding, of course, the positions of the particles
with $l_i=0$). Second, the presence of the exponential function enables us to avoid the gravitational paradox (the Neumann-Seeliger paradox). The $\overline \rho$-term
does not spoil this property because the averaged gravitational potential $\overline \varphi$ is equal to zero. Third, the gravitating masses can be distributed
completely arbitrarily. It is worth noting that, for different reasons, the arguments in favour of the open Universe were also provided in the recent paper
\cite{Barrow}.

We consider the potential \rf{2.1} against the cosmological background. In the $\Lambda$CDM model, the scale factor of the Universe reads \cite{EZcosm1}
\be{2.2}
\tilde
a=\left(\frac{\Omega_M}{\Omega_{\Lambda}}\right)^{1/3}\left[\left(1+\frac{\Omega_{\Lambda}}{\Omega_M}\right)^{1/2}
\sinh\left(\frac{3}{2}\Omega_{\Lambda}^{1/2}\tilde
t\right)+\left(\frac{\Omega_{\Lambda}}{\Omega_M}\right)^{1/2}\cosh\left(\frac{3}{2}\Omega_{\Lambda}^{1/2}\tilde t
\right)\right]^{2/3}\, ,
\ee
where we introduced dimensionless variables $\tilde a = a/a_0$, $\tilde t = H_0 t$, and $a_0$, $H_0$ are the values of the scale factor $a$ and the Hubble "constant"
$H\equiv \dot a/a\equiv (da/dt)/a$ at the present time $t=t_0$ (without loss of generality, we can put $t_0=0$). The standard density parameters are
\be{2.3} \Omega_M=\frac{\kappa\overline\rho c^4}{3H_0^2a_0^3},\quad \Omega_{\Lambda}=\frac{\Lambda c^2}{3H_0^2}\, ,
\ee
where $\kappa\equiv 8\pi G_N/c^4$ and $\Lambda$ is the cosmological constant. It is worth noting that the solution \rf{2.2} is common for any value of the curvature
parameter $\mathcal K$ because the density parameter for the curvature $|\Omega_{\mathcal K}|=|\mathcal K|c^2/(a_0^2H_0^2) \ll 1$ \cite{7WMAP} (where $\mathcal
K=-1,0,+1$ for open, flat and closed Universes, respectively), and we  can drop the spatial curvature term from the Friedmann equation. According to the seven-year WMAP
observations \cite{7WMAP}, $H_0\approx70\, \mbox{km/sec/Mpc}\approx 2.3\times 10^{-18}\mbox{sec}^{-1}\approx (13.7\times 10^9)^{-1}\mbox{yr}^{-1}$, $\Omega_M\approx0.27$
and $\Omega_{\Lambda}\approx0.73$.

Below, we shall consider astrophysical objects (galaxies and their groups) deep inside of the cell of uniformity, i.e. for physical distances $R \lesssim $ 150 Mpc. As
we have shown in \cite{EZcosm1}, the comoving distances in the cell of uniformity are much less than 1: $l_i \ll 1$. For such small distances, we can use the Cartesian
coordinates. Then, eq. \rf{2.1} reads
\be{2.4}
\varphi ({\bf r})=-G_N
\sum_{i}\frac{m_i}{|{\bf r}-{\bf r}_i|}+
\frac{4\pi G_N\overline\rho}{3}\, ,
\ee
where ${\bf r}_i$ is the comoving radius-vector of the $i$-th gravitating mass $m_i$. Its Lagrange function is \cite{EZcosm1}
\be{2.5}
\mathcal{L}_i=-\frac{m_i\varphi_i}{a}+\frac{m_i a^2 v_i^2}{2}\, ,
\ee
where
\be{2.6}
\varphi_i ({\bf r}_i)=-G_N\sum_{j\neq i}\frac{m_j}{|{\bf r}_i-{\bf r}_j|}+\frac{4\pi G_N\overline\rho}{3}
\ee
is the gravitational potential created by all the remaining masses at the point ${\bf r}={\bf r}_i$. In eq. \rf{2.5},
\be{2.7}
v_i^2=\dot x_i^2+\dot y_i^2+\dot z_i^2\, ,
\ee
and $v_i$ is the comoving peculiar velocity of the $i$-th gravitating source. It has the dimension $(time)^{-1}$. We would remind that we work in the weak-field limit
where physical peculiar velocities are much less than the speed of light: $a v_i\ll c$.

It can be easily verified that, with respect to the physical coordinates
\be{2.8}
X_i=ax_i,\quad Y_i=ay_i,\quad Z_i=az_i\, ,
\ee
the Lagrange function is
\be{2.9}
\mathcal{L}_i=-\frac{m_i\varphi_i}{a}+\frac{m_i}{2a^2}\left[\left(\dot X_i a- \dot a X_i\right)^2+\left(\dot Y_i a-\dot a Y_i\right)^2+\left(\dot Z_i a-\dot a
Z_i\right)^2\right]\, ,
\ee
where
\be{2.10}
\varphi_i=-G_Na\sum_{j\neq i}\frac{m_j}{|{\bf R}_i-{\bf R}_j|}+\frac{4\pi G_N\overline\rho}{3},\quad |{\bf R}_i-{\bf
R}_j|=\sqrt{\left(X_i-X_j\right)^2+\left(Y_i-Y_j\right)^2+\left(Z_i-Z_j\right)^2}\, ,
\ee
and the Lagrange equations for the $i$-th mass take the form
\ba{2.11}
-G_N\sum_{j\neq i}\frac{m_j\left(X_i-X_j\right)}{\left[\left(X_i-X_j\right)^2+\left(Y_i-Y_j\right)^2+\left(Z_i-Z_j\right)^2\right]^{3/2}}
&=&\frac1{a}\left(\ddot X_i a-\ddot a X_i\right)\, ,\\
-G_N\sum_{j\neq i}\frac{m_j\left(Y_i-Y_j\right)}{\left[\left(X_i-X_j\right)^2+\left(Y_i-Y_j\right)^2+\left(Z_i-Z_j\right)^2\right]^{3/2}}
&=&\frac1{a}\left(\ddot Y_i a-\ddot a Y_i\right)\label{2.12}\, , \\
-G_N\sum_{j\neq i}\frac{m_j\left(Z_i-Z_j\right)}{\left[\left(X_i-X_j\right)^2+\left(Y_i-Y_j\right)^2+\left(Z_i-Z_j\right)^2\right]^{3/2}}
&=&\frac1{a}\left(\ddot Z_i a-\ddot a Z_i\right)\label{2.13}\, .
\ea
Now, we can apply these equations to real astrophysical systems such as a group or cluster of galaxies. To illustrate this, we consider first a number of abstract
simplified examples.


\section{\label{sec:3}Illustrative examples}

\setcounter{equation}{0}

In this section, we consider simplified examples where all gravitating masses are on the same plane $Z=0$, i.e. all $Z_i=0$. Then, eqs. \rf{2.11}-\rf{2.13} take the form
\ba{3.1}
 \frac{d^2\tilde X_i}{d\tilde t^2}&=&-\frac{1}{\overline m}\sum_{j\neq i}\frac{m_j(\tilde X_i-\tilde X_j)}{[(\tilde X_i-\tilde X_j)^2+(\tilde Y_i-\tilde Y_j)^2]^{3/2}}
  + \frac{1}{\tilde a}\frac{d^2\tilde a}{d\tilde t^2}\tilde X_i\, , \quad i,j=1,\ldots ,N \, , \\
\label{3.2}
\frac{d^2\tilde Y_i}{d\tilde t^2}&=&-\frac{1}{\overline m}\sum_{j\neq i}\frac{m_j(\tilde Y_i-\tilde Y_j)}{[(\tilde X_i-\tilde X_j)^2+(\tilde Y_i-\tilde Y_j)^2]^{3/2}}
+ \frac{1}{\tilde a}\frac{d^2\tilde a}{d\tilde t^2}\tilde Y_i\, , \quad i,j=1,\ldots ,N\, ,
\ea
where $N$ is a total number of masses and we introduced the dimensionless variables
\ba{3.3}
\tilde X_i&=&X_i\left(\frac{H_0^2}{G_N\overline{m}}\right)^{1/3}= \frac{X_i}{0.95\mbox{Mpc}}\left(\frac{10^{12}M_\odot}{\overline m}\right)^{1/3}\, ,\nn \\
\tilde Y_i&=&Y_i\left(\frac{H_0^2}{G_N\overline{m}}\right)^{1/3}= \frac{Y_i}{0.95\mbox{Mpc}}\left(\frac{10^{12}M_\odot}{\overline m}\right)^{1/3}\, ,\nn \\
\tilde a&=&\frac{a}{a_0},\quad \tilde t=H_0t=\frac{t}{13.7\times 10^9\mbox{yr}}
\ea
and $\overline m=\sum_{i=1}^N m_i /N$ is the average mass of the system. Obviously, the first terms in the right hand side of eqs. \rf{3.1} and \rf{3.2} are due to the
gravitational attraction between masses  and the second terms originate from the cosmological expansion of the Universe, which is described by eq. \rf{2.2}. We consider
the stage of the accelerated expansion, i.e. $d^2\tilde a/d\tilde t^{\, 2} >0$. Therefore, the competition between these two mechanisms defines the dynamical behavior of
the masses. Depending on the initial conditions, they either collide with or move off each other. Let us demonstrate this with two particular examples.

\subsection{Three gravitating masses: $N=3$}

\begin{figure*}[htbp]
\centerline{\includegraphics[width=2.5in,height=2.5in]{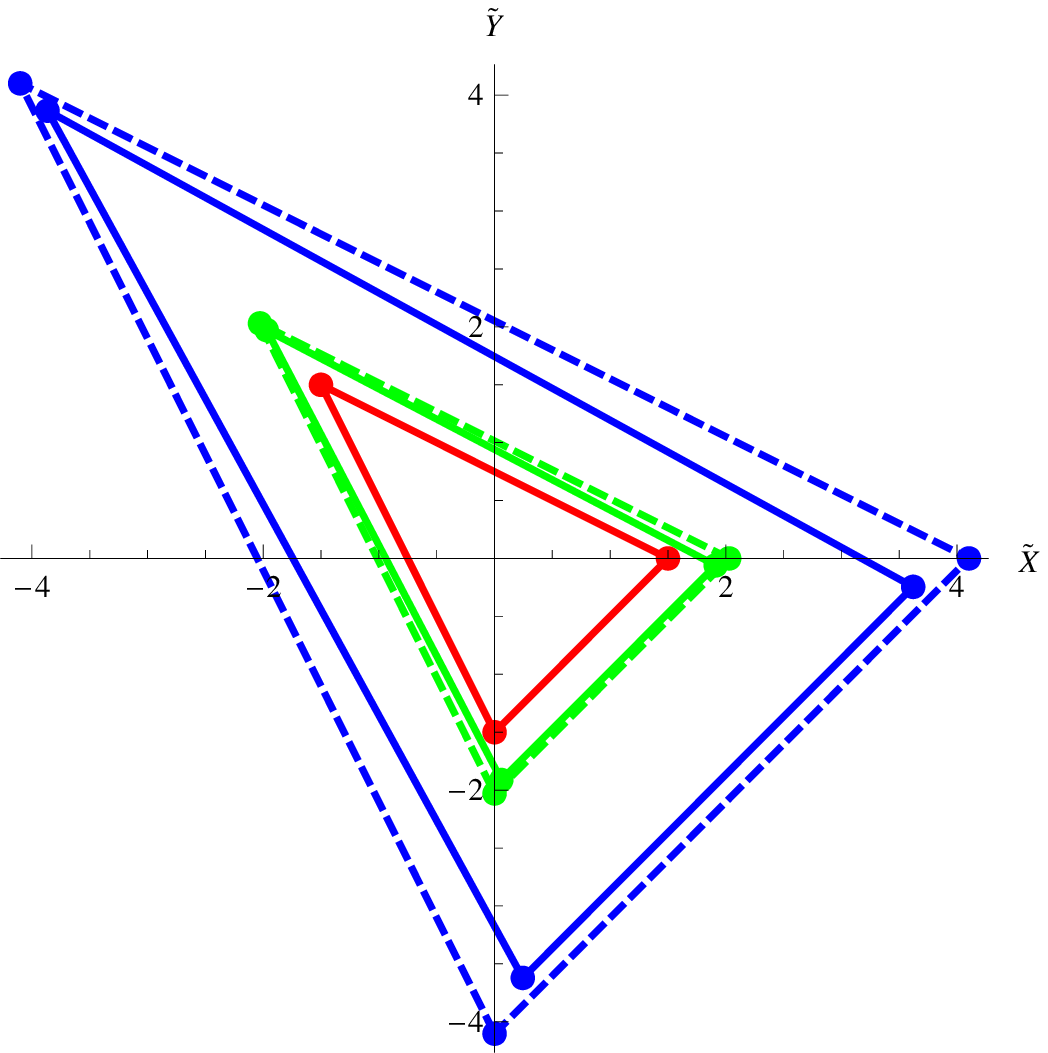}
\includegraphics[width=2.5in,height=2.5in]{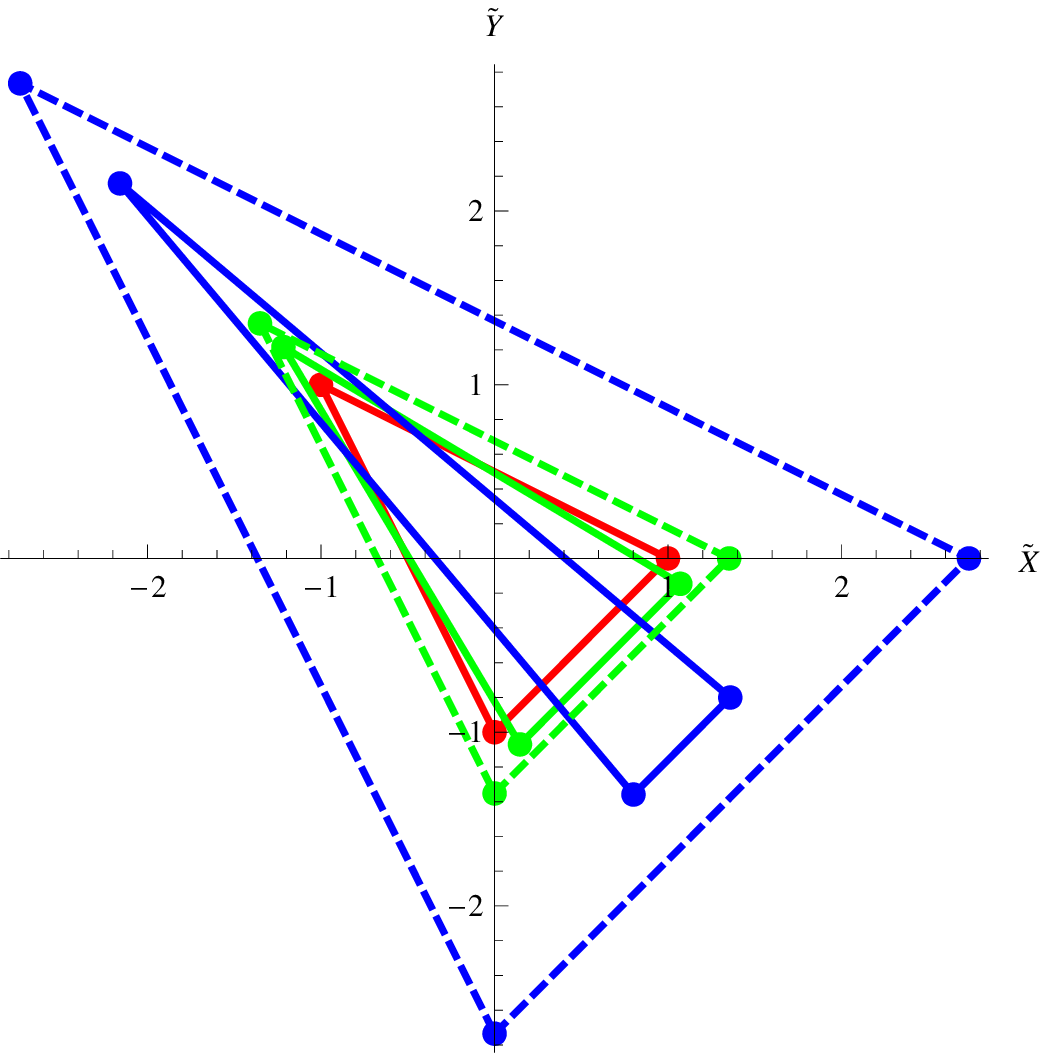}}
\caption {Dynamics of three gravitating masses with zero initial velocities. The solid red triangles describe the initial positions at $\tilde t=0$. The green and blue
triangles correspond to the positions at the moments $\tilde t=1$ and $\tilde t=2$, respectively.  Solid green and blue triangles take into account both cosmological
expansion and gravitational attraction, while the corresponding dashed triangles disregard this attraction. Depending on the initial conditions, the gravitating masses
move away from each other because the cosmological expansion prevails the gravitational attraction (the left panel), or some of masses can collide with each other in the
case of the prevalence of the attraction (the right panel). \label{triangle}}
\end{figure*}

Here, we study dynamics of three gravitating masses $(N=3)$ for two different cases. First, we consider the case with the initial coordinates $(0,-1.5)$, $(1.5,0)$ and
$(-1.5,1.5)$ for the first ($i=1$), second ($i=2$) and third ($i=3$) gravitating masses, respectively, and with zero initial velocities\footnote{The components of the
velocity are given by the formula $\dot X_i=a\dot x_i+HX_i$. Therefore, to get zero value for this expression at some moment $t_0$, the peculiar velocity should
compensate the Hubble velocity at this moment.} $d\tilde X_i/d\tilde t|_{\tilde t=0}=0,\, d\tilde Y_i/d\tilde t|_{\tilde t=0}=0,\, i=1,2,3$. For simplicity, we also take
$m_1=m_2=m_3=\overline m$. The numerical solution of eqs. \rf{3.1} and \rf{3.2} shows (see figure \ref{triangle}, the left panel) that in this case the cosmological
expansion prevails over the gravitational attraction for all three masses and all of them move off each other. The red (inner) solid triangle corresponds to the initial
positions at the initial moment $\tilde t=0$. Then, we depict green (middle) and blue (outer) triangles at the moments $\tilde t=1$ and $\tilde t=2$, respectively. Solid
triangles take into account both cosmological expansion and gravitational attraction, while the dashed triangles correspond to pure cosmological expansion. Obviously,
the dashed triangles are similar but the gravitational attraction spoils this similarity (see the solid ones). Additionally, the solid triangles are inside of the
corresponding dashed ones because the gravitational attraction slows the recession.

In the second case, the initial coordinates are $(0,-1)$, $(1,0)$ and $(-1,1)$, respectively, with the same zero initial velocities at $\tilde t=0$. Then, the results of
the numerical solutions for times $\tilde t=1$ (green triangles) and $\tilde t=2$ (blue triangles) are shown in figure \ref{triangle}, the right panel. Similar to the
previous case, solid (dashed) triangles correspond to the presence (absence) of the gravitational attraction. Therefore, the dashed triangles form the similar triangles.
The solid green and blue triangles demonstrate that the gravitational attraction between first $(i=1)$ and second $(i=2)$ masses prevail the cosmological expansion, and
these masses approach each other. However, the cosmological expansion dominates for the third $(i=3)$ mass and this mass moves away from the other two.

\subsection{Four gravitating masses: $N=4$}

\begin{figure*}[htbp]
\centerline{\includegraphics[width=2.5in,height=2.5in]{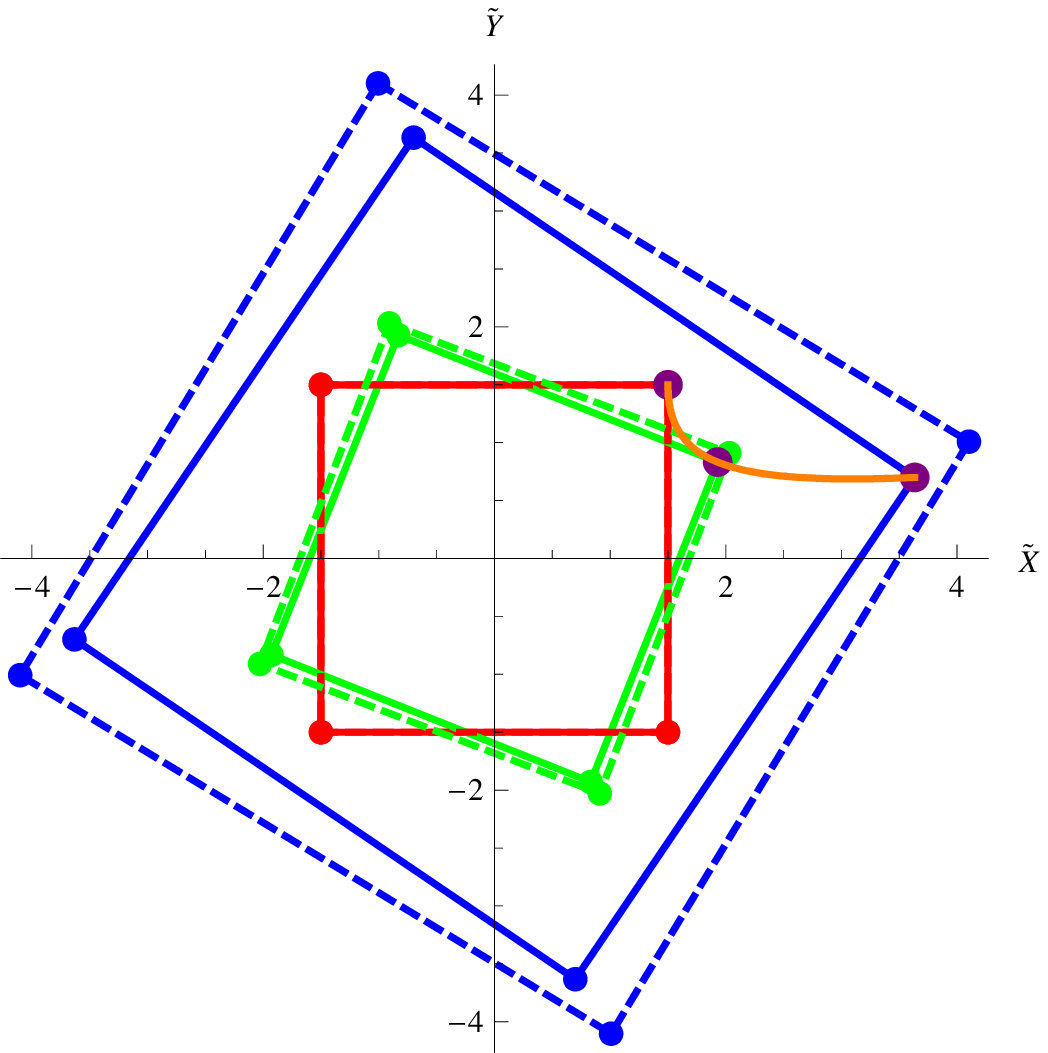}
\includegraphics[width=2.5in,height=2.5in]{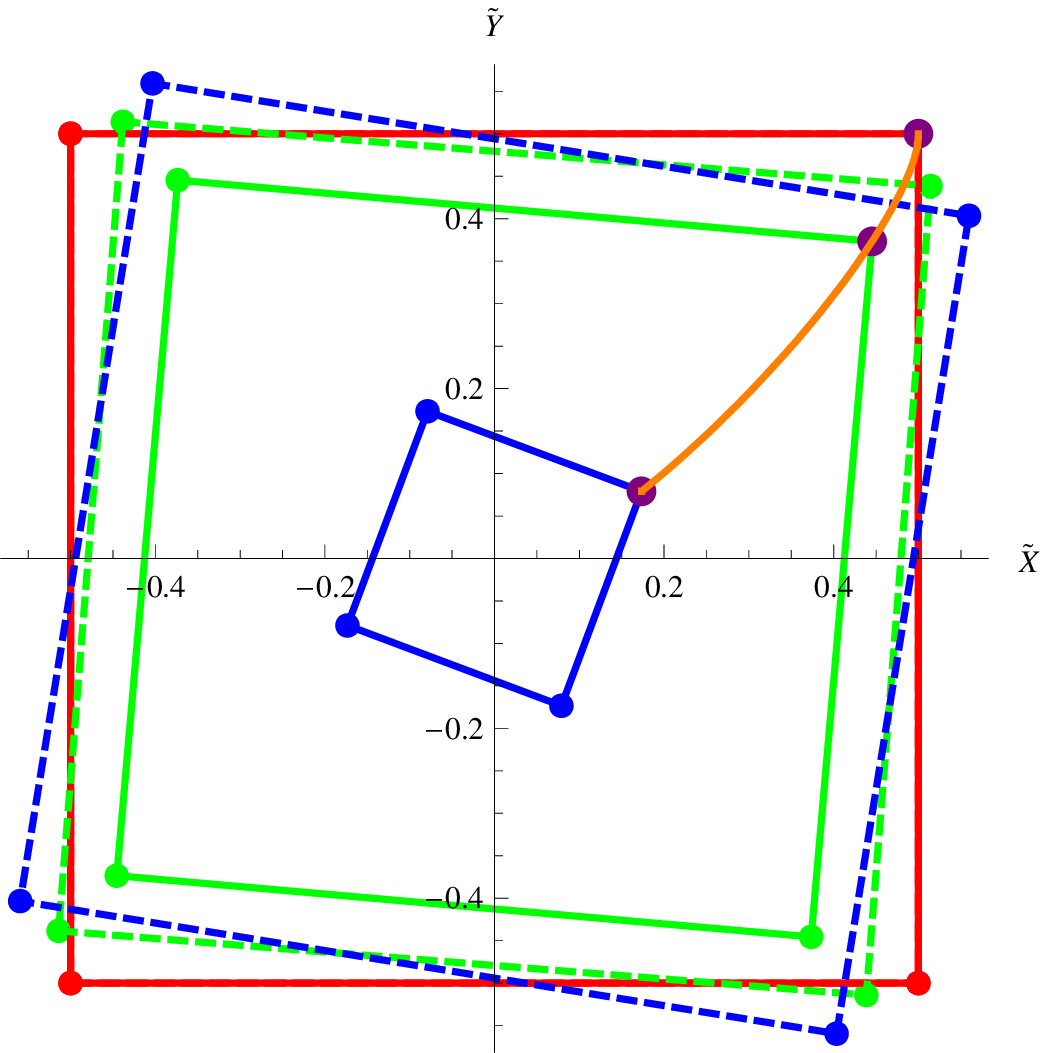}}
\caption {Dynamics of four gravitating masses with non-zero initial velocities. The solid red squares correspond to the initial positions of masses at $\tilde t=0$, and
green and blue squares show their positions at times $\tilde t=1$, $\tilde t=2$ (the left panel) and $\tilde t=0.3$, $\tilde t=0.6$ (the right panel). Solid green and
blue squares take into account both cosmological expansion and gravitational attraction, while the corresponding dashed squares disregard this attraction. The orange
curved line depicts the trajectory of one of the masses. The left panel demonstrates the situation when the cosmological expansion prevails over the gravitational
attraction and all masses move away from each other. The right panel corresponds to the opposite case when the gravitational attraction dominates and masses approach
each other. \label{square}}
\end{figure*}

Here, we study dynamical behavior of four gravitating masses $(N=4)$. The results of numerical solutions of eqs. \rf{3.1} and \rf{3.2} are depicted in figure
\ref{square}. The left panel demonstrates the situation when the cosmological expansion prevails over the gravitational attraction and all masses move away from each
other. The initial coordinates here are $(1.5,1.5)$, $(1.5,-1.5)$, $(-1.5,-1.5)$ and $(-1.5,1.5)$. The right panel corresponds to the opposite case when the
gravitational attraction dominates and masses approach each other. Here, the initial coordinates are $(0.5,0.5)$, $(0.5,-0.5)$, $(-0.5,-0.5)$ and $(-0.5,0.5)$. The
meaning of color and type of lines is the same as for the previous example, i.e. the red squares correspond to the initial positions of masses at $\tilde t=0$ and green
and blue squares show their positions at times $\tilde t=1$, $\tilde t=2$ (the left panel) and $\tilde t=0.3$, $\tilde t=0.6$ (the right panel), respectively. Solid
green and blue squares take into account both cosmological expansion and gravitational attraction, while the corresponding dashed squares disregard this attraction. In
contrast to the previous three-mass example, here, we alow the masses to rotate clockwise setting the following initial velocities $(d\tilde X_i/d\tilde t,d\tilde
Y_i/d\tilde t)_{\tilde t=0}$: $(0,-1)$, $(-1,0)$, $(0,1)$, $(1,0)$ (the left panel) and $(0,-0.25)$, $(-0.25,0)$, $(0,0.25)$, $(0.25,0)$ (the right panel), respectively.
Additionally, the orange curved line depicts the trajectory of one of the masses.

\section{\label{sec:4}Collision between Milky Way and Andromeda}

\setcounter{equation}{0}

\subsection{Free-fall approximation}

Now, we want to apply our method to real astrophysical objects. For this purpose, we consider our local group of galaxies, which consists of two giant galaxies (our
Milky Way (MW) and Andromeda (M31)) and approximately 40 dwarf galaxies.  At the present time, these giant galaxies are located at the distance 0.78 Mpc\footnote{It is
worth noting that our local group of galaxies forms the region of overdensity located inside the underdensity area. We can easily estimate the size/radius of this area
from the formula $R\sim [3M/(4\pi \bar \rho_{phys})]^{1/3}$, where $M$ is the total mass of the group and $\bar \rho_{phys}$ is the average mass density of matter in the
Universe. For our group this is a few megaparsecs, e.g., $R\sim 2.54$ Mpc for $M\approx 2.6\times 10^{12}M_{\odot}$ and $\bar \rho_{phys}\approx 0.2556\times
10^{-29}\mbox{g}/\mbox{cm}^3$. This radius can be enlarged if we include the mass of the Intra-Group Matter (IGrM). According to \cite{WhiteNelson}, IGrM can contain up
to 30\% of the group mass. In this case, for our local group $R\sim 2.77$ Mpc. Therefore, MW and M31 with their separation distance 0.78 Mpc are deep inside the
underdensity region.} and  move towards each other with the speed 120 km/sec \cite{CoxLoeb}. Therefore, in future they may encounter. The collision time was estimated
recently in the paper \cite{CoxLoeb}, where the authors used the hydrodynamic approach. They found that the average time for the first passage is 2.8 Gyr and for the
final merger is 5.4 Gyr. It is of interest to estimate also this time using our mechanical approach. We consider these two galaxies as point-like gravitating masses.
Obviously, such approach is valid at distances greater than the sizes of galaxies. For these two galaxies, we can apply our method  up to the separation distance of the
order of 100 Kpc, where the process of merger starts \cite{CoxLoeb}\footnote{According to the simulations carried out in \cite{Deason1}, this distance can be extended up
to 120-150 Kpc.}. The intergalactic/intragroup medium density is estimated as $5\div200$ times the average density of the Universe \cite{Freeland,Fang}. So, we split our
investigation  into two steps.  First, in this subsection, we neglect the Intra-Group Matter (IGrM) in our calculations. Then, in the next subsection we take into
account the dynamical friction caused by IGrM.

{}From eq. \rf{2.9}, it can be easily seen that the two-particle Lagrange function for two gravitating masses/galaxies (marked as the points A and B) is
\ba{4.1}
\mathcal{L}_{AB}&=&G_N\frac{m_A m_B}{\left\vert{\bf R}_A-{\bf R}_B\right\vert}+\frac{m_A}{2a^2}
\left[\left(\dot X_Aa-\dot a X_A\right)^2+\left(\dot Y_Aa-\dot a Y_A\right)^2+\left(\dot Z_Aa-\dot a Z_A\right)^2\right]+\nonumber\\
&+&\frac{m_B}{2a^2}\left[\left(\dot X_Ba-\dot a X_B\right)^2+\left(\dot Y_Ba-\dot a Y_B\right)^2+\left(\dot Z_Ba-\dot a Z_B\right)^2\right]\, .
\ea
Let us introduce the projections $L_X,L_Y$ and $L_Z$ of the distance between these masses and the coordinates of the center of mass:
\ba{4.2}
X_A-X_B&=& L_X\, , \\
\label{4.3}\cfrac{m_AX_A+m_BX_B}{m_A+m_B}&=&X_0\,
\ea
and the similar expressions for $L_Y,L_Z$ and $Y_0,Z_0$. Therefore, the absolute value of the distance is $\left\vert{\bf R}_A-{\bf
R}_B\right\vert=\sqrt{L_X^2+L_Y^2+L_Z^2}=L>0$. Then, eq. \rf{4.1} reads
\ba{4.4}
&{}&\mathcal{L}_{AB}=G_N\frac{m_A m_B}{L}+\frac{1}{2a^2}\left\{\frac{m_Am_B}{m_A+m_B}\left(\dot{L}_X^2 a^2-2\dot{L}_XL_X\dot a a+\dot a^2 L_X^2+\right.\right.\nonumber\\
&{}&+\left.\dot{L}_Y^2 a^2-2\dot{L}_Y L_Y\dot a a+\dot{a}^2 L_Y^2+\dot{L}_Z^2a^2-2\dot{L}_ZL_Z\dot a a+\dot a^2L_Z^2\right)+\nonumber\\
&{}&+(m_A+m_B)\left.\left[\left(\dot X_0 a- \dot a X_0\right)^2+\left(\dot Y_0 a-\dot a Y_0\right)^2+\left(\dot Z_0 a- \dot a Z_0\right)^2\right]\right\}\, ,
\ea
or, in spherical coordinates,
\ba{4.5}
&{}&\mathcal{L}_{AB}=G_N\frac{m_A m_B}{L}+\frac1{2}\frac{m_A m_B}{m_A+m_B}\left(\frac{\dot a^2}{a^2}
L^2-2\frac{\dot a}{a}\dot L L+\dot L^2+L^2\dot\theta^2+L^2\sin^2\theta\dot\psi^2\right)+\nonumber\\
&{}&+\frac1{2a^2}(m_A+m_B)\left[\left(\dot X_0 a- \dot a X_0\right)^2+\left(\dot Y_0 a-\dot a Y_0\right)^2+\left(\dot Z_0 a- \dot a Z_0\right)^2\right]\, .
\ea
It can be easily verified that $X_0$ satisfies the equation
\be{4.6}
a\ddot X_0-\ddot a X_0=0\,
\ee
with the following solution:
\be{4.7}
\dot X_0=H X_0 +\frac{a_0}{a}\left(\dot X_{0(in)}-H_0 X_{0(in)}\right)\, ,
\ee
where $a_0=a(t_0), H_0=H(t_0), X_{0(in)}=X_{0}(t_0)$ and $\dot X_{0(in)}=\dot X_{0}(t_0)$ are values at the initial time $t=t_0$.
Therefore, $\dot X_0$ satisfies asymptotically (with increasing $a$) the Hubble law. The same conclusion takes place for $\dot Y_0$ and $\dot Z_0$.

Let us investigate now the relative motion of the galaxies. For this motion, the Lagrange function is
\be{4.8}
\tilde{\mathcal{L}}_{AB}=G_N\frac{m_A m_B}{L}+\frac1{2}\frac{m_A m_B}{m_A+ m_B}\left(\frac{\dot a^2}{a^2}L^2-2\frac{\dot a}{a}\dot L L+\dot
L^2+L^2\dot\psi^2\right)\, ,
\ee
where without loss of generality we put $\theta=\pi/2$. Therefore, the Lagrange equation for the separation distance is
\be{4.9}
\ddot L=-G_N\frac{2\overline{m}}{L^2}+\frac{M^2}{\mu^2 L^3}+\frac{\ddot a}{a}L\, ,
\ee
where we introduced the reduced mass, the average mass and the angular momentum:
\be{4.10}
\frac{m_Am_B}{m_A+m_B}\equiv\mu\, ,\quad \overline{m}=\frac{m_A+m_B}{2}\, ,\quad \mu L^2\dot\psi\equiv M=\mbox{const}\, .
\ee
The first term in the right hand side of \rf{4.9} is due to the gravitational attraction, the second term is the centrifugal force and the third term originates from the
cosmological expansion of the Universe. To integrate eq. \rf{4.9}, we rewrite it with respect to the dimensionless quantities similar to ones in \rf{3.3}:
\ba{4.11}
\tilde L&=&L\left(\frac{H_0^2}{G_N\overline{m}}\right)^{1/3}\approx \frac{L}{0.95\mbox{Mpc}}\left(\frac{10^{12}M_\odot}{\overline m}\right)^{1/3}\, ,\nn \\
\tilde V&=&V\left(\frac{1}{H_0G_N\overline{m}}\right)^{1/3}\approx \frac{V}{67\mbox{km/sec}}\left(\frac{10^{12}M_\odot}{\overline m}\right)^{1/3}\, , \label{4.11}\\
\tilde M&=&M\frac{\overline{m}}{\mu}\left(\frac{H_0}{G_N^2\overline{m}^5}\right)^{1/3}\approx \frac{L}{0.95\mbox{Mpc}}\;
\frac{V_{\perp}}{67\mbox{km/sec}}\left(\frac{10^{12}M_\odot}{\overline m}\right)^{2/3}\, ,\nn \ea
where the radial velocity $V= \dot L\to \tilde V = d\tilde L/d\tilde t$ and the transverse velocity  $V_{\perp}= L\dot \psi$. Then, eq. \rf{4.9} reads
\be{4.12}
\frac{d^2\tilde L}{d\tilde t^2}=-\frac{2}{\tilde L^2} + \frac{\tilde M^2}{\tilde L^3}+\frac{1}{\tilde a}\frac{d^2\tilde a}{d\tilde t^2}\tilde L\, .
\ee

Now, we integrate eq. \rf{4.12} for parameters corresponding to the galaxies MW and M31. The masses of MW and M31 are of the order of $10^{12}M_\odot$ and $1.6\times
10^{12}M_\odot$ respectively \cite{CoxLoeb}\footnote{We take these values because we want to compare our results with the conclusions of this paper. More recent
publications indicate both a little bit higher values of the mass of MW \cite{McMillan,Boylan} and lower values \cite{Deason}.}. The separation distance at present
time\footnote{Without loss of generality we may put $t_0=0,\ \tilde t_0=0$.} $t=t_0$ is $L_0\approx 0.78\,\mbox{Mpc}\to \tilde L_0\approx 0.753$, and the galaxies
approach each other with the radial velocity $V_0\approx -120\,\mbox{km/sec}\to \tilde V_0 \approx -1.633$ \cite{CoxLoeb}.

\begin{figure*}[htbp]
\centerline{\includegraphics[width=3.0in,height=2.5in]{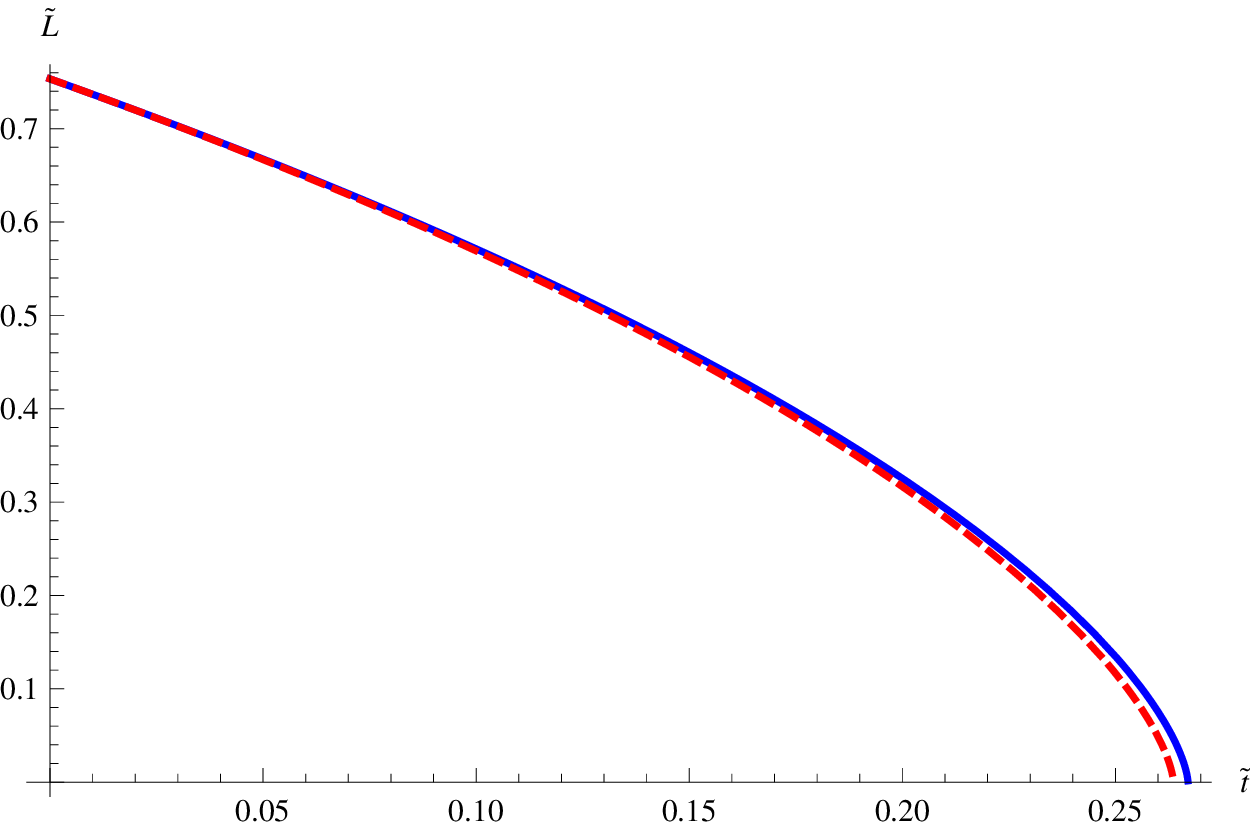}
\includegraphics[width=3.0in,height=2.5in]{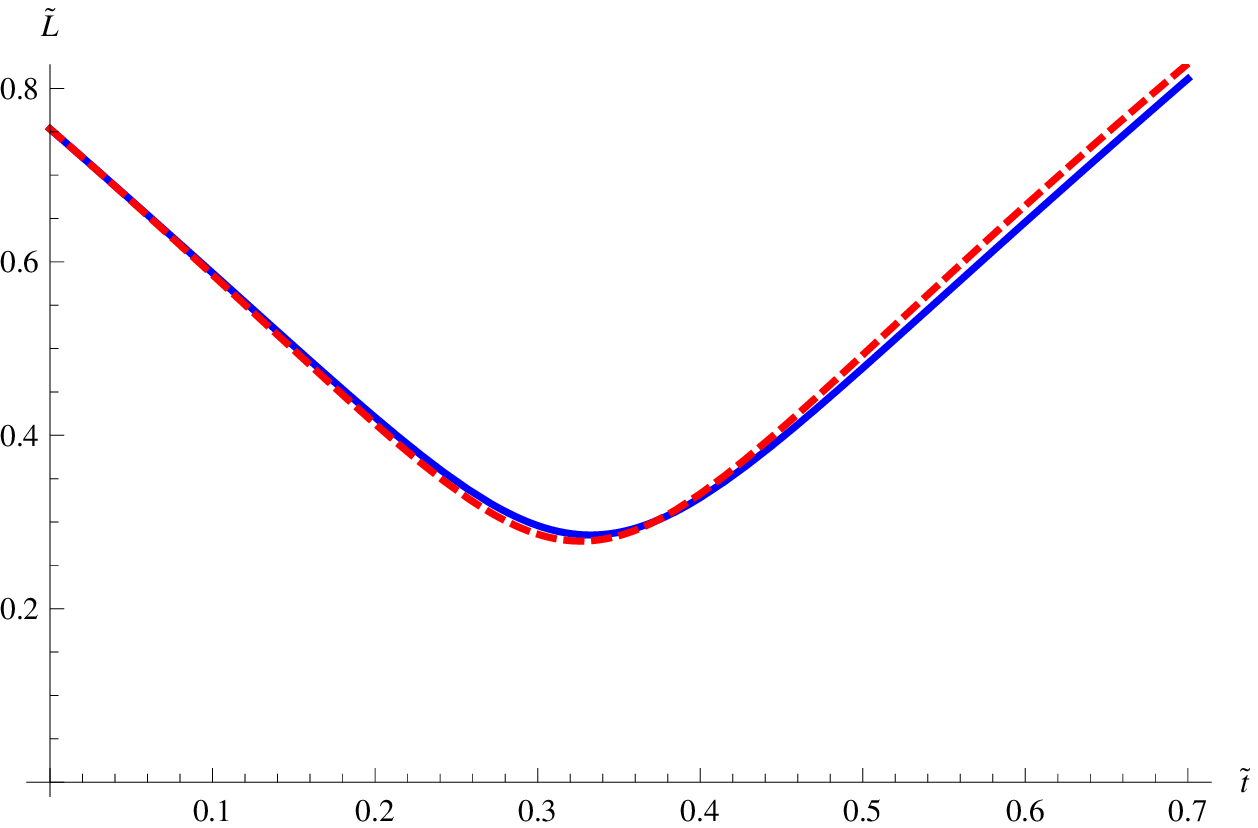}}
\caption {These figures show the change with time of the separation distance between the Milky Way and Andromeda starting from the present ($\tilde t=0$) in the case of
the absence of the dynamical friction. The initial (i.e. at present time) separation distance and radial relative velocity are $0.78$ Mpc and $-120$ km/sec, respectively
($0.753$ and $-1.633$ in dimensionless units). The solid blue lines take into account both gravitational attraction and cosmological expansion, while the red dashed
lines disregard the cosmological expansion. The transverse velocity is absent in the left panel. Here, the collision between galaxies takes palace in $\tilde t \approx
0.267\to t\approx 3.68$ Gyr from present (the blue line). In the right panel, the transverse velocity is equal to 100 km/sec. Here, the collision is absent and the
smallest separation distance is $\tilde L \approx 0.28 \to L\approx 290$ Kpc at time $\tilde t \approx 0.324 \to t\approx 4.44$ Gyr from present. For both pictures, the
effect of the cosmological expansion is very small for the considered period of time. \label{andromeda}}
\end{figure*}

First, let us consider briefly the case of the zero angular momentum $M=0$. If $\ddot a>0$, as it happens at the present stage of the Universe evolution, we can
introduce a distance of zero acceleration $L_{cr}$ where $\ddot L=0$:
\be{4.13}
\tilde L_{cr}(\tilde t)=\left(\frac{2}{-q(\tilde t)}\right)^{1/3}\, ,
\ee
where the deceleration parameter $q=-(1/H^2)(\ddot a/a)=-(d^2\tilde a/d\tilde t^2)/\tilde a$. In the $\Lambda$CDM model, we get from the Friedmann equations that at
present time $q_0\approx \Omega_M/2 -\Omega_{\Lambda} \approx -0.595$, and the zero acceleration distance is $\tilde L_{cr}(\tilde t_0)\approx 1.5$. If at $t=t_0$ the
relative velocity $\dot L(t_0)=0$, then gravitating masses run away from each other (collide with each other) in the future for $L(t_0)>L_{cr}(t_0)$
($L(t_0)<L_{cr}(t_0)$). Obviously, the separation distance $\tilde L_0 \approx 0.753$ between MW and Andromeda is less than $\tilde L_{cr}(\tilde t_0)$. Additionally,
they have the non-zero radial velocity towards each other. Therefore, they will collide with each other.  The result of numerical solution of this collision for $M=0$
is shown in figure \ref{andromeda}, the left panel. The solid blue line takes into account both gravitational attraction and cosmological expansion while the red dashed
line disregards the cosmological expansion. This picture demonstrates that the effect of the expansion is very small for relative motion of MW and M31. For example, the
time to collision (from present) is $\tilde t\approx 0.2670\to 3.68$Gyr and $\tilde t\approx 0.2636\to 3.63$Gyr for blue and red lines, respectively. We remind that our
approach works up to the separation distance $\tilde L\approx 0.1 \to 100$ Kpc when the stage of the galaxy merger starts.

Let us turn now to the case of the non-zero angular momentum. The observations indicate the proper motion of Andromeda perpendicular to our line of sight. This
transverse velocity $V_{\perp 0}=V_{\perp}(t_0)$ is less than 200 km/sec \cite{Peebles}. In  \cite{Loeb}, the authors found an even smaller estimate: $V_{\perp 0} \sim
100$ km/sec. In our calculations, we will adhere to this value, and with the help of eq. \rf{4.11} we get $\tilde M \approx 1.029$. It can be easily seen from eq.
\rf{4.12} that fall to the center is absent because of the centrifugal barrier ($M\neq 0$). Therefore, the collision of the galaxies is possible if the smallest
separation distance between them (which corresponds to the turning point) is less than the merger distance $100-150$ Kpc. The result of the numerical integration of eq.
\rf{4.12} is shown in figure \ref{andromeda}, the right panel. The solid blue line takes into account both gravitational attraction and cosmological expansion, while the
red dashed line disregards the cosmological expansion. Similar to the previous case, the effect of the cosmological expansion is very small for the considered period of
time. This picture demonstrates that for the given transverse velocity, the smallest separation distance is $\tilde L \approx 0.28 \to 290$ Kpc at time $\tilde t \approx
0.324 \to 4.44$ Gyr from present. This distance is much bigger than the merger distance. Therefore, for the chosen initial conditions, the collision between the Milky
Way and Andromeda is absent. The collision may take place for a smaller transverse velocity. For example, if $V_{\perp 0} \approx 60$ km/sec, then the smallest
separation distance is 100 Kpc, which can be sufficient to start the merger.

\subsection{Dynamical friction}

Let us take now into account the Intra-Group Matter (IGrM). It is well known that a massive body with a mass $M$ moving through surrounding matter, which consists of
discrete particles of the mass $m$, will lose its momentum and kinetic energy due to gravitational interaction with these particles. Such effect is called dynamical
friction. The force of the dynamical friction is given by the Chandrasekhar formula \cite{Binney}:
\be{4.14} \frac{d{\bf V}_M}{dt} =- \frac{4\pi Q \; G_N^2 M\rho_{ph,m}}{V^3_M}\left[\mbox{erf}(\chi) -\frac{2\chi}{\sqrt{\pi}}\exp\left(-\chi^2\right)\right]{\bf V}_M\, ,
\ee
where ${\bf V}_M$ is the physical velocity of the mass $M$, $\rho_{ph,m}$ is the physical rest mass density of IGrM, $\chi \equiv V_M/(\sqrt{2} \sigma)$ and erf is the
error function. Here, $Q \equiv (1/2) \ln\left(1+\lambda^2\right)$ is the so called Coulomb logarithm defined by the largest impact parameter $b_{max}$, the initial
relative velocity $V_0$ and the masses $M$ and $m$: $\lambda = b_{max} V_0^2/[G_N(M+m)]\approx b_{max} V_0^2/(G_N M)$. The formula \rf{4.14} is defined with respect to a
frame where the IGrM particles have the Maxwell's speed distribution with the dispersion $\sigma = \sqrt{kT/m}$. The typical value of the IGrM temperature in the Local
Group is \cite{CoxLoeb} $T\sim 10^5 \mbox{K} \to kT \sim 8.6$ eV.

Therefore, the Milky Way and Andromeda should slow down moving through the IGrM because of the dynamical friction \rf{4.14}. Then, equations \rf{3.1} and \rf{3.2}
describing the dynamics of the galaxies MW and M31 (labelled as A and B, respectively) are modified as follows:
\ba{4.15}
&{}&\frac{d^2\tilde X_i}{d\tilde t^2}=-\frac{1}{\overline m}\frac{m_j(\tilde X_i-\tilde X_j)}{[(\tilde X_i-\tilde X_j)^2+(\tilde Y_i-\tilde Y_j)^2]^{3/2}}+
\frac{1}{\tilde a}\frac{d^2\tilde a}{d\tilde t^2}\tilde X_i  \\
&{}&- \frac{3\, Q\, m_i\, \alpha}{2\, \overline{m}\, \tilde v_{pec,i}^3}\left[\mathrm{erf}\left(\tilde\chi_i\right)- \frac{2\tilde \chi_i}{\sqrt{\pi}}
\exp\left(-\tilde\chi_i^2\right)\right] \left(\frac{d \tilde X_i}{d\tilde t}-\frac{1}{\tilde a}\frac{d\tilde a}{d\tilde t}\tilde X_i\right)\, ,\quad i,j=A,B;\; i\neq j\, ,\nn\ea
\ba{4.17}
&{}&\frac{d^2\tilde Y_i}{d\tilde t^2}=-\frac{1}{\overline m}\frac{m_j(\tilde Y_i-\tilde Y_j)}{[(\tilde X_i-\tilde X_j)^2+(\tilde Y_i-\tilde Y_j)^2]^{3/2}}+
\frac{1}{\tilde a}\frac{d^2\tilde a}{d\tilde t^2}\tilde Y_i  \\
&{}&- \frac{3\, Q\, m_i\, \alpha}{2\, \overline{m}\, \tilde v_{pec,i}^3}\left[\mathrm{erf}\left(\tilde\chi_i\right)- \frac{2\tilde \chi_i}{\sqrt{\pi}}
\exp\left(-\tilde\chi_i^2\right)\right] \left(\frac{d \tilde Y_i}{d\tilde t}-\frac{1}{\tilde a}\frac{d\tilde a}{d\tilde t}\tilde Y_i\right)\, ,\quad i,j=A,B;\; i\neq j\, , \nn \ea
where we assume that the IGrM particles have the Maxwell's speed distribution in a frame comoving with the Hubble flow. In this case ${\bf V}_M$ in \rf{4.14} is the
peculiar velocity of MW and M31:
%
\ba{4.19} \tilde{{\bf v}}_{pec,i}&=&\left(\frac{d\tilde X_i}{d\tilde t}-\frac{1}{\tilde a}\frac{d\tilde a}{d\tilde t}\tilde X_i,\frac{d\tilde Y_i}{d\tilde t}-
\frac{1}{\tilde a}\frac{d\tilde a}{d\tilde t}\tilde Y_i\right)\, , \\
\label{ 4.20} \tilde v_{pec,i}&=&\left[\left(\frac{d\tilde X_i}{d\tilde t}-\frac{1}{\tilde a}\frac{d\tilde a}{d\tilde t}\tilde X_i\right)^2+
\left(\frac{d\tilde Y_i}{d\tilde t}-\frac{1}{\tilde a}\frac{d\tilde a}{d\tilde t}\tilde Y_i\right)^2\right]^{1/2}\, ,\quad i=A,B\, .
\ea
As in equations \rf{3.3} and \rf{4.11}, tilde denotes dimensionless quantities. Additionally, $\tilde \chi_i =\tilde v_{pec,i}/(\sqrt{2}\tilde \sigma)$ and $\tilde
\sigma=(\sigma/H_0)\left[H_0^2/(G_N\overline{m})\right]^{1/3}$. Regarding the physical  rest mass density of IGrM, we define it in the terms of the critical density:
$\rho_{ph,m}=\alpha\rho_{cr}=\alpha \left(3H_0^2\right)/(8\pi G_N)$. For the IGrM in the Local Group, we shall take $\alpha \sim 10$ \cite{CoxLoeb}. To estimate the
Coulomb logarithm $Q$ for the Local Group, first, we should take into account that the matter density $\rho_{ph,m}$ begins to decrease at scales where the cosmological
expansion starts to dominate over the gravitational attraction, i.e. approximately at 1 Mpc. Here, the parameter $\alpha \sim 5$  \cite{CoxLoeb}. Therefore, $b_{max}
\sim 1$ Mpc. Then, taking for $V_0$ the typical peculiar velocity 100 km/sec and for $M$ the value $10^{12}M_\odot$, we get that $Q\sim 1$.

\begin{figure*}[htbp]
\centerline{\includegraphics[width=3.0in,height=2.5in]{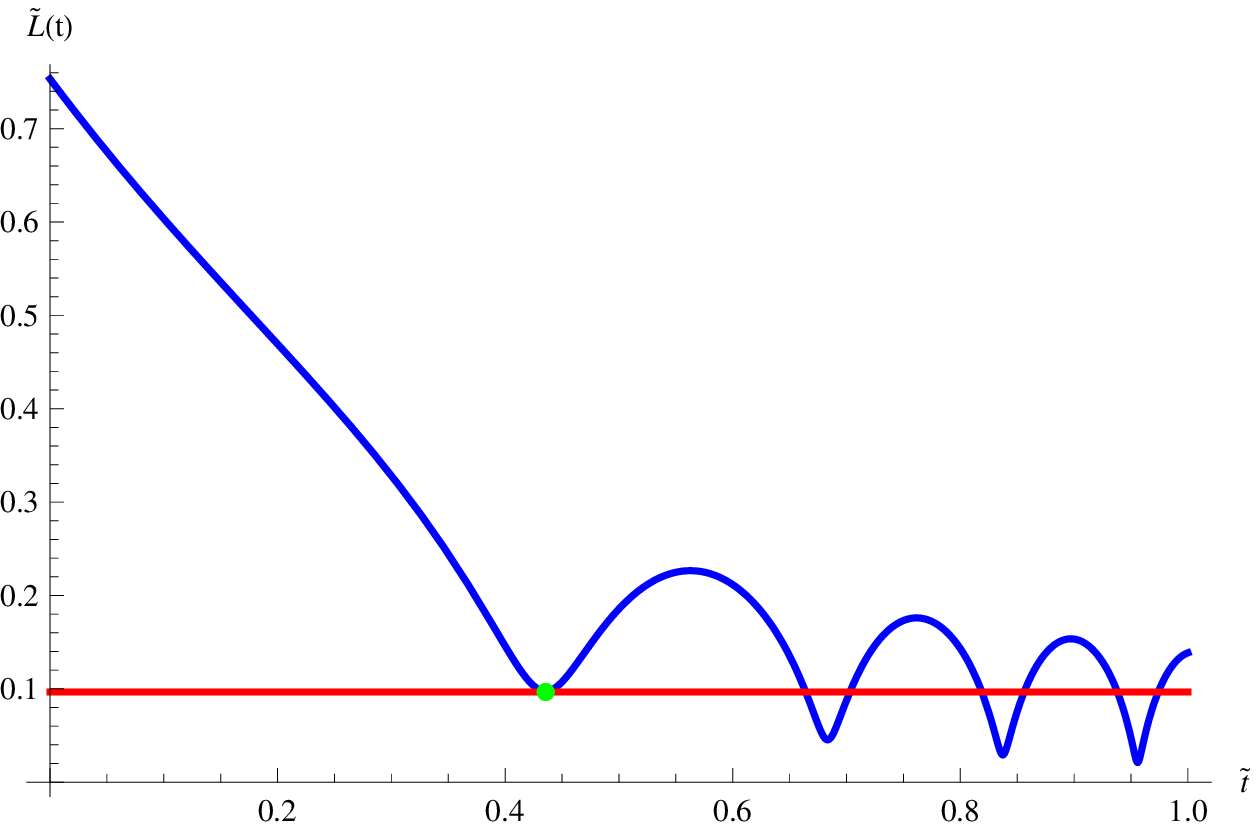}
\includegraphics[width=3.0in,height=1.3in]{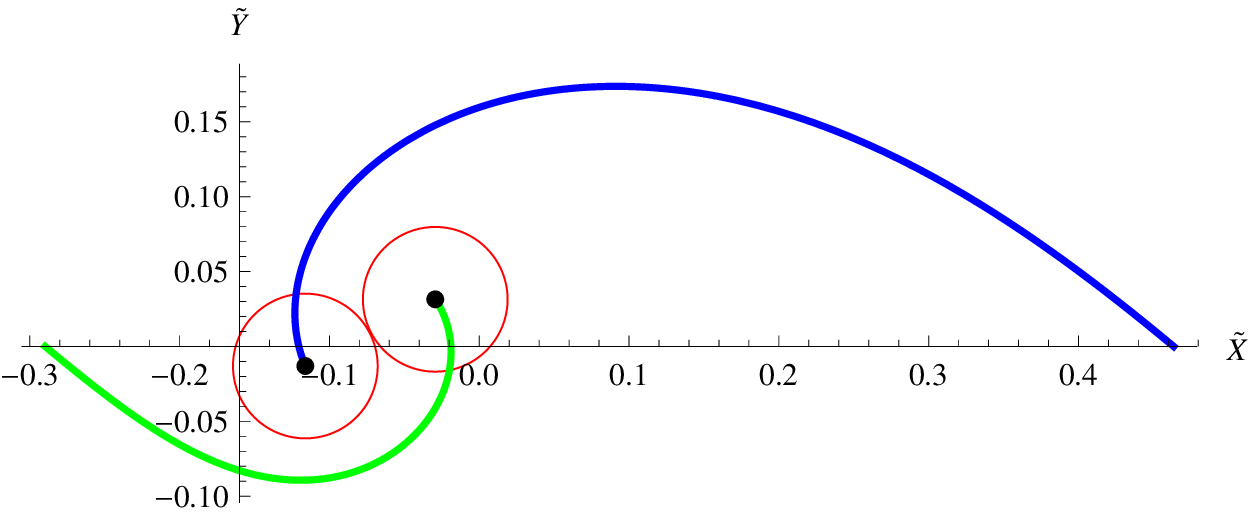}}
\caption {These figures show the change with time of the separation distance between the Milky Way and Andromeda
(the left panel) and the corresponding trajectories for the MW (the blue line) and M31 (the green line) in the right panel in the case of dynamical friction. The initial
conditions are chosen as in the right panel of the figure \ref{andromeda}.  The dynamical friction is calculated for the dispersion parameter $\tilde\sigma
=\tilde\sigma_1 = 1.17$. For this value of $\tilde\sigma$, the first close passage occurs at the separation distance $L=100$ Kpc (see the green point in the left panel
and two red touched circles in the right panel) which corresponds to the touch of the galaxies. For this and smaller values of $\tilde\sigma$,
the touch of the galaxies will take place during the first passage.\label{graphmin}}
\end{figure*}

\begin{figure*}[htbp]
\centerline{\includegraphics[width=3.0in,height=2.5in]{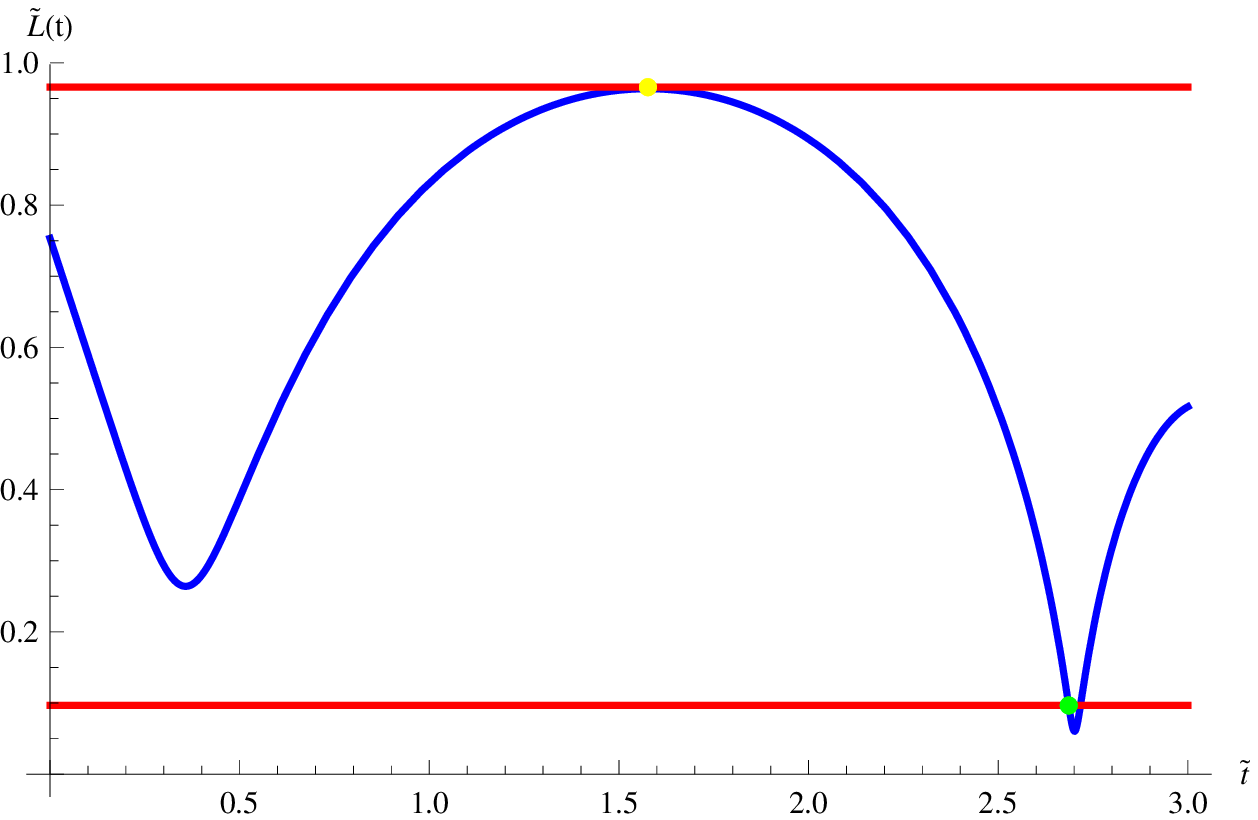}
\includegraphics[width=3.0in,height=3.1in]{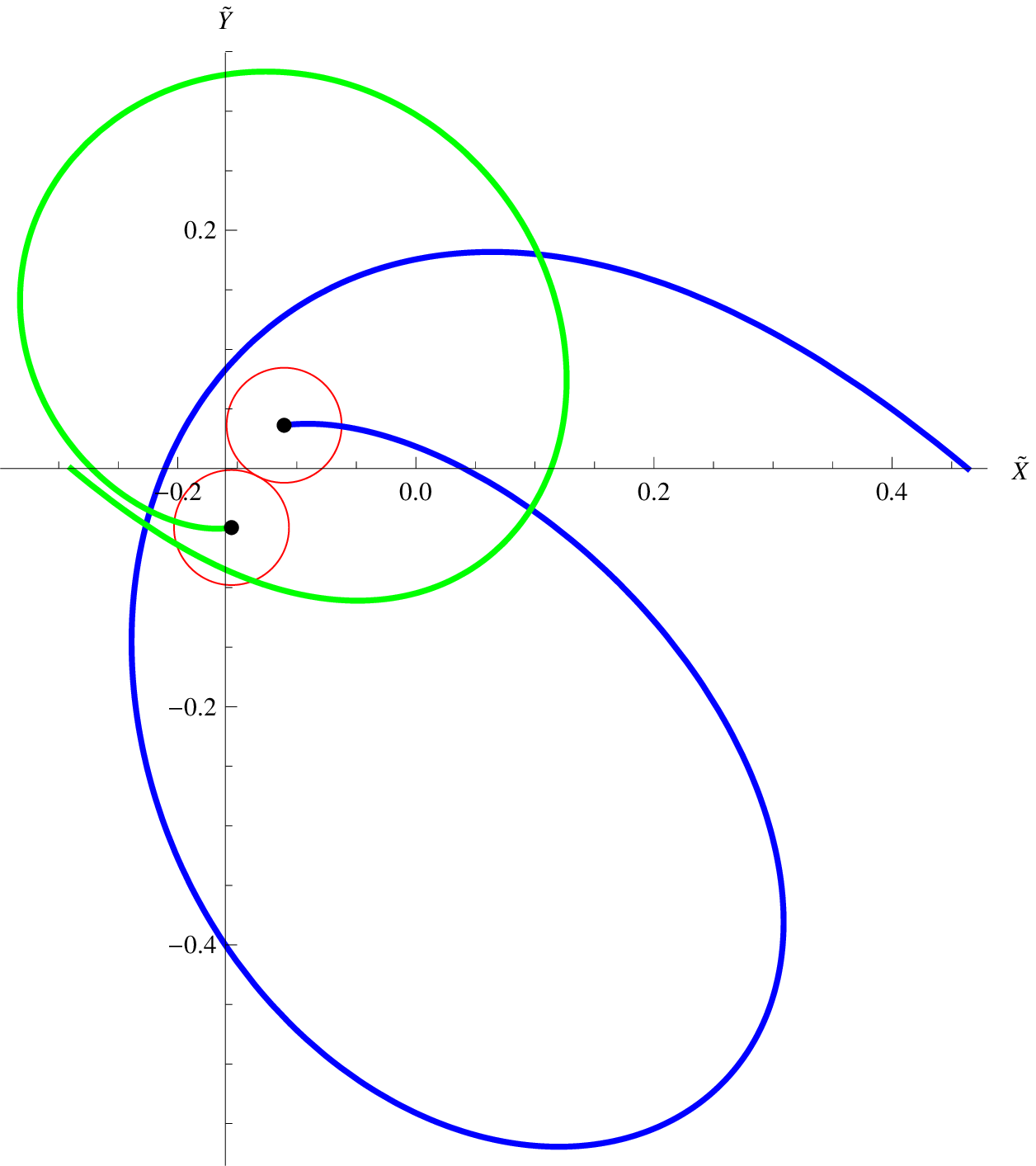}}
\caption {These figures are drawn in the case of the dynamical friction with the dispersion parameter $\tilde\sigma =\tilde\sigma_2 = 2.306$. For this value of
$\tilde\sigma$, there is no touch of the galaxies during the first passage because the closest separation distance here is larger than 100 Kpc (the bottom red line in
the left panel). After that, the galaxies grow apart to the turning point at the separation distance 1 Mpc from each other (see the yellow point on the upper red line in
the left panel). At these and greater distances, the rest mass density of IGrM decreases and the dragging effect of the dynamical friction can be too small to force the
galaxies to converge again. Therefore, for $\tilde \sigma > \tilde\sigma_2$, the merger of the galaxies becomes problematic. \label{graphmax}}
\end{figure*}

Let $X_{i}(t)$ and $Y_{i}(t)$ be the barycentric coordinates of the MW ($i=A$) and M31 ($i=B$), that is the origin of coordinates is in the center of mass of MW and M31.
In this case, the initial values of the coordinates and velocities are: $X_A(t_0)=m_B L_0/(m_A+m_B)$, $X_B(t_0)=-m_A L_0/(m_A+m_B)$, $Y_A(t_0)=0$, $Y_B(t_0)=0$ and $\dot
X_A(t_0)=m_B V_0/(m_A+m_B)$, $\dot X_B(t_0)=-m_A V_0/(m_A+m_B)$, $\dot Y_A(t_0)= m_B V_{\perp 0} /(m_A+m_B)$, $\dot Y_B(t_0)= -m_A V_{\perp 0} /(m_A+m_B)$, where we use
the notations from the previous subsection. Now, taking the values of the parameters from the previous subsection (the case of the nonzero angular momentum, e.g.,
$V_0=-120$ km/sec and $V_{\perp 0}=100$ km/sec), we can integrate numerically the equations \rf{4.15} and \rf{4.17}. We take into account both the cosmological expansion and
the gravitational attraction, although, as we have seen above, the influence of the expansion is not significant within the scales of interest $L \leq 1$ Mpc. The result
of these calculations is depicted in figures \ref{graphmin} and \ref{graphmax} where the left panels show the change in time of the separation distance and the right
panels describe the trajectories of the galaxies. These figures demonstrate that, for fixed values of masses of galaxies, their initial conditions and the parameter
$\alpha$, the relative dynamical behavior depends on the dispersion parameter $\sigma = \sqrt{kT/m}$ which in turn is defined by the ratio of the temperature $T$ and the
mass $m$ of the particles of IGrM. Comparatively little is known about truly intergalactic medium. Most probably this is a mixture of the baryonic matter (mainly in the
form of ionized hydrogen) and dark matter. There is great variety of candidates for dark matter with masses ranging from $\mu$eV$\div$eV (e.g., axions) to TeV (e.g.,
WIMPs). Therefore, in the formula for the dispersion $\sigma$, the parameters $T$ and $m$ are some effective values. It makes sense not to specify them separately, but
to consider their ratio, i.e. $\sigma^2$. As we mentioned above, our approach works up to the first touch of the galaxies which occurs approximately at the separation
distance 100 Kpc between their centers. In the left panels, this event is marked by the green points on the bottom red lines. In this case, the merger of the galaxies
will take place. Our approach does not describe this process. The continuations of the lines (the separation distance) after the first touch is very schematic. In the
right panels, this event corresponds to the touch of two red circles. The distance between their centers is equal to 100 Kpc. We do not continue the trajectories after
this first touch.

We found two characteristic values for the dimensionless parameter $\tilde \sigma$. The first one is $\tilde \sigma_1 = 1.17$ and corresponds to the situation when the
first close passage occurs at the separation distance $L=100$ Kpc (see the green point in the left panel and two red touched circles in the right panel in the figure
\ref{graphmin}) which corresponds to the touch of the galaxies. Obviously, for all $\tilde \sigma < \tilde \sigma_1$, this distance will be less than 100 Kpc and the
first touch of the galaxies will take place during the first passage. For the bigger values of $\tilde \sigma$, the first passage occurs at the separation distance
larger than 100 Kpc. The second characteristic value is $\tilde \sigma_2 = 2.306$ and describes the situation when the galaxies, after the first close passage, grow
apart to the turning point at the separation distance 1 Mpc from each other (see the yellow point on the upper red line in figure \ref{graphmax}, the left panel). At
these and greater distances, the rest mass density of IGrM decreases and the dragging effect of the dynamical friction can be too small to force the galaxies to converge
again. Therefore, for $\tilde \sigma > \tilde\sigma_2$, the merger of the galaxies becomes problematic. For $\tilde \sigma_1 < \tilde \sigma < \tilde \sigma_2$, the
 touch will take place during the second passage.  It is of interest to estimate masses of the IGrM particles which correspond to these characteristic values of
$\tilde \sigma$. The masses $m$ can be expressed via the temperature $T$ and dimensionless dispersion $\tilde \sigma$ as follows: $m (\mbox{MeV})\approx
\left\{\left[kT(\mbox{erg})/8.464\times 10^{13}(\mbox{cm}^2/\mbox{sec}^2)\right]\times 0.5604\times 10^{27}(\mbox{MeV}/\mbox{g})\right\}/\tilde\sigma^2$. The temperature
of IGrM in the Local Group is usually estimated as $T\sim 10^5$ K \cite{CoxLoeb}. Then, for this value of $T$, we get $m_1 \sim 67$ MeV and $m_2 \sim 17$ MeV for $\tilde
\sigma_1$ and $\tilde \sigma_2$, respectively. Therefore, for the chosen initial conditions and the value of $T$, the touch of the galaxies will take place during the
first passage
for the IGrM particle masses $m \geq 67$ MeV and the merger can be problematic for masses lighter than $17$ MeV.

\section{\label{sec:5}Formation of Hubble flows in the vicinity of the Local Group}

\setcounter{equation}{0}

To study the formation of the Hubble flows in the vicinity of our group of galaxies, we need to determine the spatial distribution of vectors of acceleration of
astrophysical objects (e.g., dwarf galaxies) in the gravitational field of two giant galaxies taking into account the cosmological expansion of the Universe. Obviously,
near the galaxies, the vector must be oriented in the direction of galaxies due to the gravitational attraction, and with the distance from galaxies he has to turn in
the opposite direction due to the cosmological accelerated recession.

Let us investigate this effect for our Local Group. For this purpose, we consider a test particle/dwarf galaxy
in the gravitational field of Andromeda and Milky Way. We study the picture at present time when the separation distance between M31 and MW is $L_0=0.78$ Mpc, and we do
not take into account the relative motion of these galaxies. Of course, we can also consider dynamical evolution of this system but this effect is out of the scope of
this section. It can be easily seen from the Lagrange function \rf{2.9} that the Lagrange equation for a test particle is
\be{5.1}
\frac{d}{dt}\left({\bf V}-\frac{\dot a}{a}{\bf R}\right)=-\frac{1}{a}\frac{\partial\varphi}{\partial{\bf R}}+\frac{\dot a^2}{a^2}{\bf R}-\frac{\dot a}{a}{\bf
V}
\ee
or, equivalently,
\be{5.2}
\dot {\bf V}-\frac{\ddot a}{a}{\bf R} =-\frac{1}{a}\frac{\partial\varphi}{\partial{\bf R}}\, .
\ee
If the gravitational field is absent (i.e. $\varphi \equiv 0$) then this equation has the following solution:
\be{5.3} {\bf V}=\frac{\dot a}{a}{\bf R}+\frac{{\bf const}}{a}\, , \ee
where the first term is the Hubble velocity and the second term is the peculiar velocity.
In dimensionless variables (see \rf{3.3} and \rf{4.11}) and without peculiar velocity this solution reads
\be{5.4}
\tilde {\bf V} = \frac{1}{\tilde a}\frac{d\tilde a}{d\tilde t}\tilde{\bf R}=\frac{H}{H_0}\tilde{\bf R}\, .
\ee
In the case of our Local Group, the gravitational potential is (see eq. \rf{2.10}):
\be{5.5}
\varphi=\varphi_A+\varphi_B,\quad \varphi_A=-aG_N\frac{m_A}{\left\vert{\bf R}_A-{\bf R}\right\vert},\quad \varphi_B=-aG_N\frac{m_B}{\left\vert{\bf R}_B-{\bf
R}\right\vert}\, ,
\ee
where, similar to the previous section, we mark MW and M31 by letters A and B, respectively. Here, we omitted the constant term $\sim \overline \rho$ because it does not
contribute to equations of motion.

For numerical solution, we rewrite eq. \rf{5.2} in dimensionless variables \rf{3.3} and \rf{4.11} as
\be{5.6} \tilde {\bf W}=\frac{d\tilde {\bf V}}{d\tilde t}=\frac{1}{\tilde a}\frac{d^2\tilde a}{d\tilde t^2}\tilde {\bf R}-\frac{1}{\tilde
a}\frac{\partial\tilde\varphi}{\partial \tilde {\bf R}}\, , \ee
where we introduced additionally the dimensionless potential
\be{5.7}
\tilde \varphi=\frac{1}{a_0\left(H_0 G_N\overline{m}\right)^{2/3}}\varphi\, .
\ee
It makes sense to rewrite \rf{5.6} in components. For example, for the $X$-component we have
\be{5.8} \tilde W_x=\frac{d\tilde V_x}{d\tilde t}=\frac{1}{\tilde a}\frac{d^2\tilde a}{d\tilde t^2}\tilde X-\frac{1}{\tilde a}\frac{\partial\tilde\varphi}{\partial
\tilde X}\ee
with
\ba{5.9}
\frac{1}{\tilde a}\frac{\partial \tilde\varphi}{\partial\tilde X}&=&\frac{m_A}{\overline{m}}\frac{\tilde X-\tilde X_A}{\left[\left(\tilde X-\tilde
X_A\right)^2+\left(\tilde Y-\tilde Y_A\right)^2+\left(\tilde Z-\tilde Z_A\right)^2\right]^{3/2}}\nn \\
&+&\frac{m_B}{\overline{m}}\frac{\tilde X-\tilde X_B}{\left[\left(\tilde
X-\tilde X_B\right)^2+\left(\tilde Y-\tilde Y_B\right)^2+\left(\tilde Z-\tilde Z_B\right)^2\right]^{3/2}}
\ea
and similar for the $Y$- and $Z$-components. The absolute value of the dimensionless acceleration is
\be{5.10} |\tilde {\bf W}|=\left|\frac{d\tilde {\bf V}}{d\tilde t}\right|=\frac{1}{(H_0^4G_N\overline{m})^{1/3}}\left|\frac{d {\bf V}}{d t}\right|=\sqrt{\left(\tilde
W_x\right)^2+\left(\tilde W_y\right)^2+\left(\tilde W_z\right)^2}\, . \ee

Eqs. \rf{5.6} and \rf{5.8} demonstrate how the cosmological expansion competes with the gravitational attraction. There is a characteristic region defined by the
condition
\be{5.11}
\left|\frac{1}{\tilde a}\frac{d^2\tilde a}{d\tilde t^2}\tilde {\bf R}\right| \sim \left|\frac{1}{\tilde a}\frac{\partial\tilde\varphi}{\partial \tilde {\bf
R}}\right|\, ,
\ee
where the gravitational attraction is balanced by the cosmological expansion. The gravitational attraction is stronger inside of this region (for smaller distances) and
the cosmological expansion prevails this attraction outside of this area (for larger distances).
It takes place both for positive and negative values of the cosmological acceleration $\ddot a$.\footnote{Here, we mean just the absolute values of accelerations.
Obviously, in the decelerated Universe, both the acceleration caused by the gravitational attraction and the acceleration due to the cosmological expansion are negative
and they never compensate each other. Nevertheless, there is a region where the absolute value of the cosmological acceleration becomes bigger than the absolute value of
the gravitational one. In this region, if the initial velocity ${\bf V}_0$ of a test object (e.g., a dwarf galaxy) is equal to the Hubble velocity $H{\bf R}_0$, then
this object will continue to follow the Hubble flow. On the other hand, if its initial velocity ${\bf V}_0$ with respect to an observer in the origin is equal to zero
(i.e. its peculiar velocity is equal to minus Hubble velocity, see footnote 2), then this test object will approach the origin after being released due to the total
negative acceleration. This will happen at any separation distance between the test object and the observer (see the corresponding discussion in \cite{dec1,dec2,dec3}).}
In the case of the accelerated expansion $\ddot a >0$, we have $|{\bf W}| \approx 0$ and this characteristic area is called the region of zero acceleration. Obviously,
the Hubble flows are formed outside of this area.

For our Local Group consisting of two giant galaxies MW and M31, we choose the origin of coordinates in the barycenter of these galaxies and $X-$axis along the line
connecting MW and M31. Therefore, $X_A=L_0 m_B/(m_A+m_B),\, X_B=-L_0 m_A/(m_A+m_B)$, and $Y_A=Z_A=0$, $Y_B=Z_B=0$. Additionally, due to the rotational symmetry around
the $X-$axis, it is sufficient to consider the plane $Z=0$.  The 3-D picture can be easily reconstructed by rotation around this axis. Therefore, we investigate the
distribution of the test body acceleration in the plane $Z=0$. For the masses MW and M31, we take values from the previous section: $m_A\approx 10^{12}M_\odot$ and
$m_B\approx 1.6\times 10^{12}M_\odot$.

\begin{figure*}[htbp]
\centerline{\includegraphics[width=4.0in,height=3in]{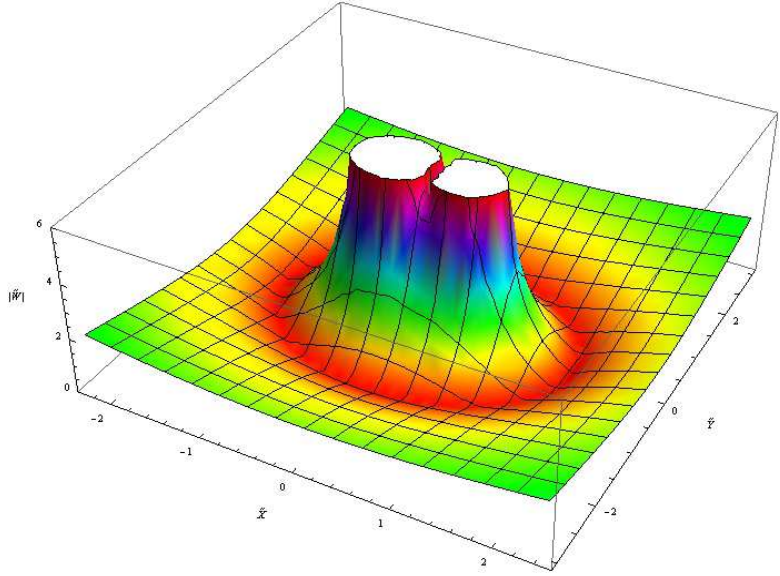}}
\caption {This plot shows the absolute value of the acceleration of dwarf galaxies in the Local Group. The red area around the peaks corresponds approximately to the
zero acceleration region. \label{camel}}
\end{figure*}

\begin{figure*}[htbp]
\centerline{\includegraphics[width=3.0in,height=3.0in]{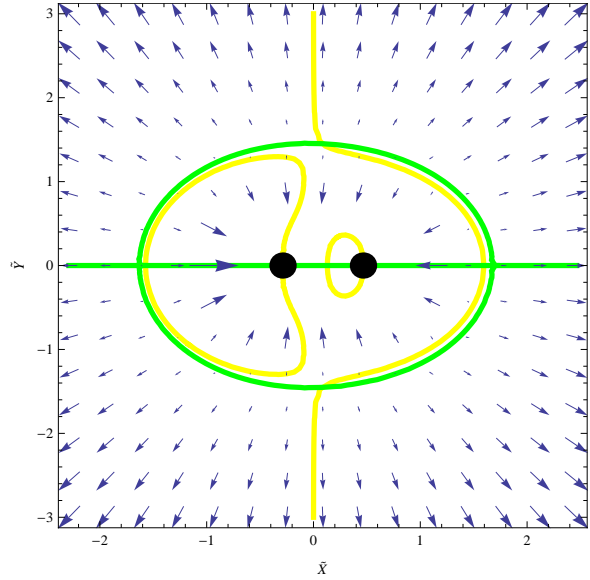}} \caption {This figure shows the vector field of the dwarf galaxy acceleration $\tilde {\bf W}$.
These vectors are directed towards the MW and M31 (the black points) in the vicinity of the galaxies and turn out with the distance from the galaxies. The yellow and
green lines correspond to the conditions $\tilde W_x=0$ and $\tilde W_y=0$, respectively. \label{hedgehog}}
\end{figure*}

On figure \ref{camel}, we depict the absolute value of the acceleration \rf{5.10} of the test body in the plane $Z=0$. This modulus decreases from large values near the
positions of MW and M31 to nearly zero (a red area around the peaks), and then it begins to increase again with the distance from the barycenter. The red area describes
approximately the region of zero acceleration. Figure \ref{hedgehog} depicts the vector field of the acceleration \rf{5.6}. It demonstrates the turn of these vectors
from the directions towards the MW and M31 (in the vicinity of these galaxies) to outside (with distance from the galaxies). The central region (near MW and M31) is
empty because we cut off the vectors with the magnitude $|\tilde {\bf W}|>3$. The yellow and green lines correspond to the conditions $\tilde W_x=0$ and $\tilde W_y=0$,
respectively. This figure shows that the vectors change their directions in the vicinity of the region $|\tilde {\bf W}|\approx 0$.

\begin{figure*}[htbp]
\centerline{\includegraphics[width=2.95in,height=2.65in]{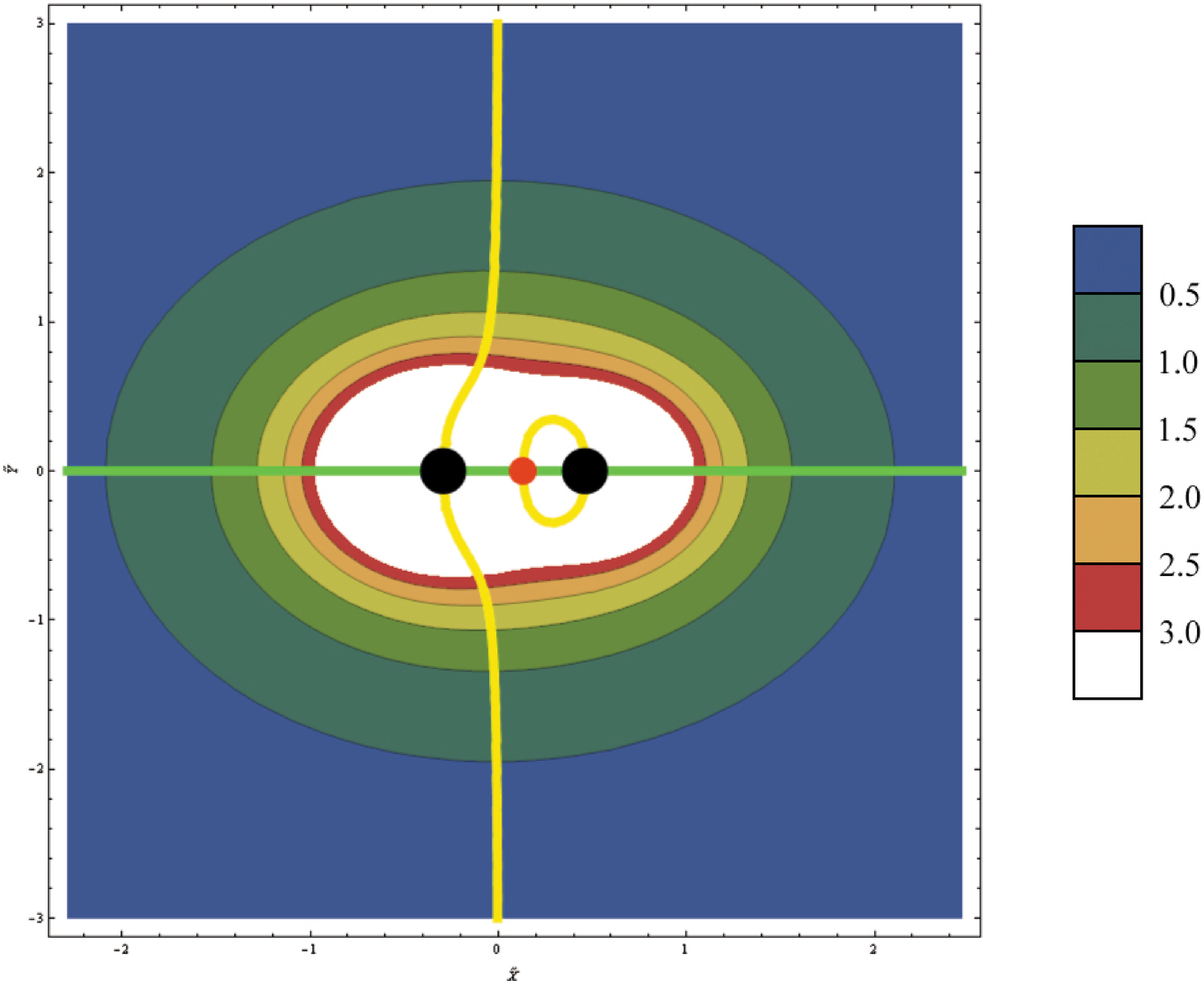}
\includegraphics[width=3.0in,height=2.7in]{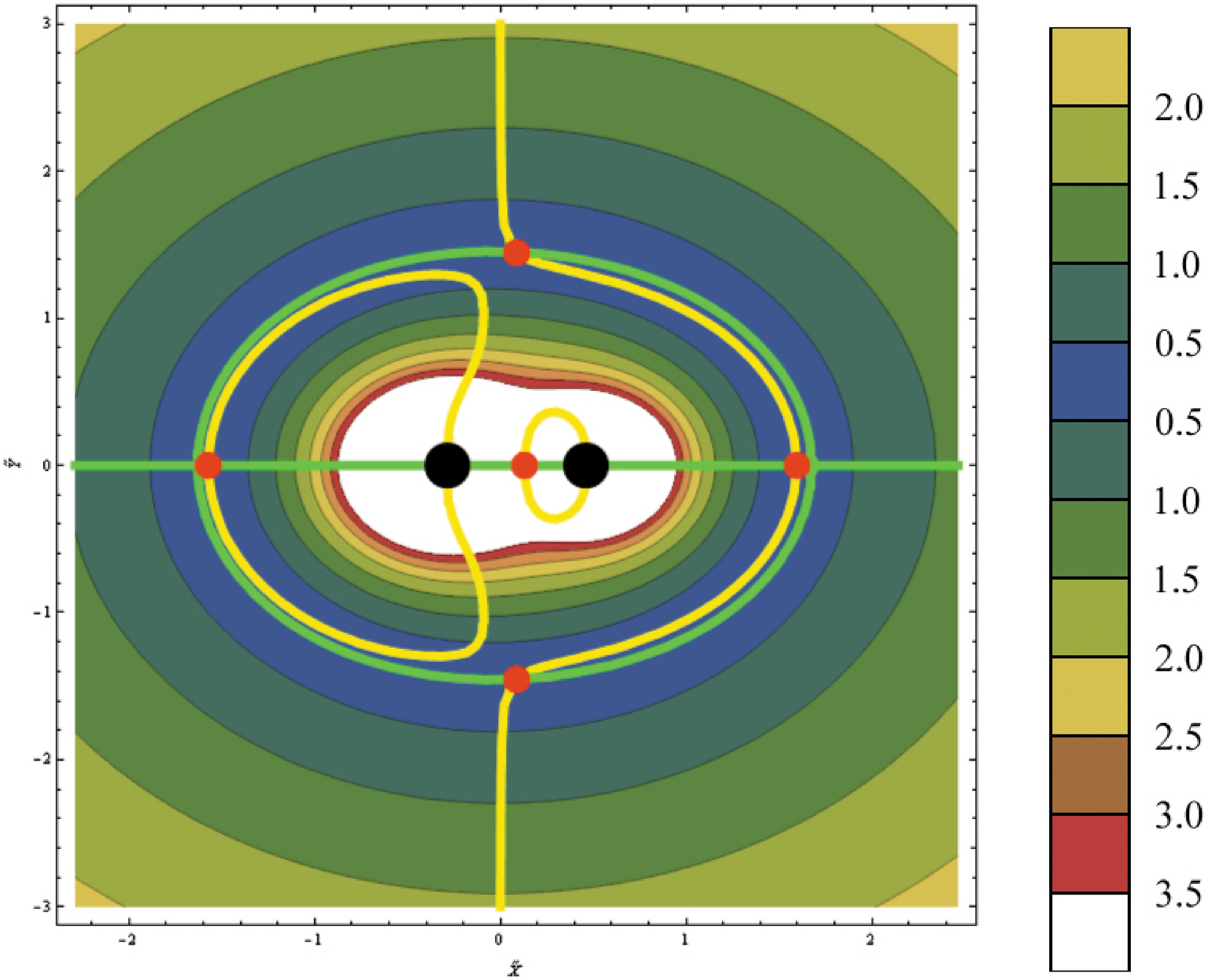}}
\caption {Here, we depict the contour plot of the absolute value of the acceleration $|\tilde {\bf W}|$ in our Local Group (left and right black points are M31 and MW,
respectively). The yellow and green lines correspond to the conditions $W_x=0$ and $W_y=0$, respectively. Red points define the positions of the zero acceleration:
$W_x=W_y=0 \to |W|=0$. The right panel takes into account both the gravitational attraction and the cosmological expansion, while the left panel disregards the
cosmological expansion. The elliptic-like green curve together with the neighboring yellow curve defines the region where $|\tilde {\bf W}|\approx 0$ (the right panel).
The vertical bars show the correspondence between the contour plot color and the absolute value of the acceleration.
\label{Lagrangepoints}}
\end{figure*}

To define more exactly the structure of the zero-acceleration surface, we draw figure \ref{Lagrangepoints}. The yellow and green lines correspond to the conditions
$W_x=0$ and $W_y=0$, respectively. Black points define the positions of the Milky Way (the right point) and Andromeda (the left point). Red points are defined by the
condition $W_x=W_y=0 \to |{\bf W}|=0$. The right panel takes into account both the gravitational attraction and the cosmological expansion, while the left panel
disregards the cosmological expansion. Obviously, in the case of only the gravitational attraction (the left panel), we have only one zero acceleration point between MW
and M31 which is the analog of the Lagrange point $L_1$. Much more reach picture happens in the presence of the cosmological accelerated expansion which competes with
the gravitational attraction (the right panel). However, this panel shows that, strictly speaking, the zero acceleration surface is absent. Here, we have two additional
red points on the $X-$axis and two vertical points. Clearly, due to the rotational symmetry, the latter two points are just the section of a zero acceleration circle by
the plane $Z=0$. Nevertheless, we can speak about the approximate zero acceleration surface because the elliptic-like green and yellow lines are very close to each other
and they define the region where $|{\bf W}|\approx 0$. We can see also that there are two regions where this surface has a discontinuity. It is clear from the rotational
symmetry that these two regions belong to the round belt (they are the section of this belt by the plane $Z=0$). Therefore, inside of the approximate zero acceleration
surface the gravitational attraction is stronger than the cosmological expansion while outside of this surface the cosmological expansion prevails over the gravitational
attraction. Obviously, the Hubble flows are formed in the latter region. Additionally, we can see that there is an asymmetry in directions along $X$ and $Y$ axes. The
characteristic distances from the barycenter to the zero acceleration surface are $|\tilde X|\approx 1.6 \to |X|\approx 1.65$ Mpc and $|\tilde Y|\approx 1.45 \to
|Y|\approx 1.5$ Mpc.
As it follows from figures \ref{hedgehog} and \ref{Lagrangepoints}, the distance $R$ where the cosmological accelerated expansion begins to prevail over the
gravitational attraction is approximately $R\approx 1.6$ Mpc, that is of the order of the scale of our local group. In other words, the cosmological constant is
significant on these scales. It is worth noting that in this section we did not take into account the IGrM in the Local Group. Obviously, the inclusion of this matter
into consideration will lead to a slight increase of the characteristic distance $R$ to the zero acceleration surface.

\section{Conclusion}

In this paper, we have considered the motion of astrophysical objects deep inside of the cell of uniformity where both the gravitational attraction between them and the
cosmological expansion of the Universe play the role. To describe this, we obtained the general system of equations of motion for arbitrary distributed inhomogeneities
in the open Universe. To show the competitive effects of gravitational attraction and cosmological recession, we considered two illustrative abstract examples where the
systems of galaxies consist of three and four galaxies.

Then, we investigated our Local Group which consists of two giant galaxies the Milky Way and Andromeda and approximately 40 dwarfs galaxies. According to recent
observations, these giant galaxies move towards each other with the relative velocity $\sim 100$ km/sec and may collide in future. Such process was investigated, e.g.,
in \cite{CoxLoeb} where the authors used the hydrodynamic approach. They found that the average time for the first passage is 2.8 Gyr and for the final merger is 5.4
Gyr. In our paper, we distinguished two different models. For the first one, we did not take into account the influence of the Intra-Group Matter. In this case, our
mechanical approach has shown that for currently known parameters of this system, the collision is hardly plausible in future because of the angular momentum. These
galaxies will reach the minimum distance of about 290 Kpc in 4.44 Gyr from present, and then begin to run away irreversibly from each other. For the second model, we
took into account the dynamical friction due to the IGrM. We found a characteristic value of the IGrM particle velocity dispersion $\tilde \sigma = 2.306$. For $\tilde
\sigma \leq 2.306$, the merger will take place, but for bigger values of $\tilde\sigma$ the merger can be problematic because the galaxies approach a region where the
dragging effect of the dynamical friction can be too small to force the galaxies to converge.If the temperature of the IGrM particles is $10^5$ K, then this
characteristic value of $\tilde \sigma$ corresponds to the IGrM particle mass 17 MeV. Therefore, for lighter masses the merger is problematic. We also have shown that
for the chosen initial conditions and the value of $T$, the touch of the galaxies will take place during the first passage of MW and M31 for the IGrM particle masses $m
\geq 67$ MeV.

We have also defined the region in the vicinity of our Local Group where the Hubble flows start to form. For such processes, the zero acceleration surface (where the
gravitational attraction is balanced by the cosmological accelerated expansion) plays the crucial role. We have taken into account that two giant galaxies MW and M31 are
located at the distance of 0.78 Mpc from each other. Obviously, if this surface exists, it does not have a spherical shape for given geometry. We have shown that such
surface is absent for the Local Group. Instead, we have found two points and one circle with zero acceleration. Nevertheless, there is the nearly closed area around the
MW and M31 where the absolute value of the acceleration is approximately equal to zero. The Hubble flows are formed outside of this area.

After finishing this article, we became aware of a recent paper \cite{new3}, which also considers the collision between the Milky Way and Andromeda. This paper is based
on the authors' measurements of a proper motion of the galaxy M31 \cite{new1}. According to these measurements, the authors found in \cite{new2} the radial and
transversal velocity of M31 with respect to the Milky Way. They are $V_{rad,M31} \equiv V_0 = -109.3$ km/sec and $V_{tan,M31} \equiv V_{\perp 0}= 17.0$ km/sec,
respectively. These values are less than those we used for our simulation. It is clear, that if the transverse velocity is so small, the merger of the galaxies MW and
M31 is inevitable, even without dynamical friction. As far as we can judge from the references cited in our paper, there is wide observational data spread for our Local
Group. The advantage of our mechanical approach lies in the fact that our equations enable us to calculate easily the dynamical evolution of the Local Group for any set
of the initial conditions.



\acknowledgments

This work was supported in part by the "Cosmomicrophysics-2" programme of the Physics and Astronomy Division of the National Academy of Sciences of Ukraine. We want to
thank the referee for his/her comments which have considerably improved the motivation of our approach and the presentation of the results. A.Zh. acknowledges the Theory
Division of CERN for the hospitality during the final stage of the preparation of this paper.



\begin{thebibliography}{99}



\bibitem{Sand1}
A. Sandage, {\em Bias properties of extragalactic distance indicators. VIII. $H_0$ from distance-limited luminosity class and morphological type-specific luminosity
functions for Sb, Sbc, and Sc galaxies calibrated using Cepheids, Astrophys. J.} {\bf 527} (1999) 479.
\bibitem{Kar et al}
I.D. Karachentsev et al, {\em The very local Hubble flow, Astronomy and Astrophysics} {\bf 389} (2002) 812  [astro-ph/0204507].
\bibitem{Kar2003}
I.D. Karachentsev, A.D. Chernin and P. Teerikorpi, {\em The Hubble flow: why does the cosmological expansion preserve its kinematical identity from a few Mpc distance to
the observation horizon?, Astrophysics} {\bf 46} (2003) 399  [astro-ph/0304250].
\bibitem{Sand2}
A. Sandage et al,
{\em The Hubble constant: a summary of the HST program for the luminosity calibration of
type Ia supernovae by means of Cepheids, Astrophys. J.} {\bf 653} (2006) 843  [astro-ph/0603647].
\bibitem{Kar2008}
I.D. Karachentsev, O.G. Kashibadze, D.I. Makarov and R.B. Tully, {\em The Hubble flow around the Local Group, MNRAS} {\bf 393} (2009) 1265  [astro-ph/0811.4610].
\bibitem{Kar2012}
I.D. Karachentsev, {\em Missing dark matter in the local Universe, Astrophysical Bulletin.} {\bf 67} (2012) 123  [astro-ph/1204.3377].
\bibitem{Chernin 2012}
A.D. Chernin et al,
{\em Dark energy in six nearby galaxy flows: synthetic phase diagrams and self-similarity, Astronomy Reports.} {\bf 56} (2012)  653.
\bibitem{CoxLoeb}
T.J. Cox and A. Loeb, {\em The collision between the Milky Way and Andromeda, MNRAS.} {\bf 386} (2008) 461  [astro-ph/0705.1170].
\bibitem{Labini}
F.S. Labini, {\em Inhomogeneities in the universe, Class. Quantum Grav.} {\bf 28} (2011) 164003
[astro-ph/1103.5974].
\bibitem{EZcosm1}
M. Eingorn and A. Zhuk,  {\em Hubble flows and gravitational potentials in observable Universe, JCAP.} {\bf 09} (2012) 026
[astro-ph/1205.2384].
\bibitem{Barrow}
J.D. Barrow, C.G. Tsagas and K. Yamamoto, {\em Do intergalactic magnetic fields imply an open Universe? Phys. Rev. D.} {\bf 86} (2012) 107302  [gr-qc/1210.1183].
\bibitem{7WMAP}
E. Komatsu et al., {\em Seven-Year Wilkinson Microwave Anisotropy Probe (WMAP) observations: cosmological interpretation, Astrophys. J. Suppl.} {\bf 192} (2011) 18
[astro-ph/1001.4538].
\bibitem{WhiteNelson}
N.J. White and A.H. Nelson, {\em Simulation of the formation of compact groups of galaxies, 12th Kingston Meeting on Theoretical Astrophysics: Computational
Astrophysics, ASP Conference Series} {\bf 123} (1997) 207; D.A. Clark and J. West (eds).
\bibitem{Deason1}
A.J. Deason, V. Belokurov, N.W. Evans and K.V. Johnston, {\em Broken and unbroken: the Milky Way and M31 stellar haloes,} [astro-ph/1210.4929].
\bibitem{Freeland}
E. Freeland, R.F. Cardoso and E. Wilcots, {\em Bent-double radio sources as probes of intergalactic gas, Astrophys. J.} {\bf 685} (2008) 858  [astro-ph/0806.3971].
\bibitem{Fang}
T. Fang et al., {\em Confirmation of X-ray absorption by WHIM in the Sculptor Wall, Astrophys. J.} {\bf 714} (2010) 1715  [astro-ph/1001.3692].
\bibitem{McMillan}
P.J. McMillan, {\em Mass models of the Milky Way, MNRAS.} {\bf 414} (2011) 2446  [astro-ph/1102.4340].
\bibitem{Boylan}
M. Boylan-Kolchin, {\em The space motion of Leo I: the mass of the Milky Way's dark matter halo,} [astro-ph/1210.6046].
\bibitem{Deason}
A.J. Deason et al., {\em The cold veil of the Milky Way stellar halo, MNRAS} {\bf 425} (2012) 2840  [astro-ph/1205.6203].
\bibitem{Peebles}
P.J.E. Peebles, S.D. Phelps, E.J. Shaya and R.B. Tully, {\em Radial and transverse velocities of nearby galaxies, Astrophys. J.} {\bf 554} (2001) 104
[astro-ph/0010480].
\bibitem{Loeb}
A. Loeb, M.J. Reid, A. Brunthaler and H. Falcke, {\em Constraints on the proper motion of the Andromeda Galaxy based on the survival of its satellite M33, Astrophys. J.}
{\bf 633} (2005) 894  [astro-ph/0506609].
\bibitem{Binney}
J. Binney and S. Tremaine, {\em Galactic Dynamics,} Princeton University Press, Princeton, New Jersey (1987).
\bibitem{dec1}
T.M. Davis, C.H. Lineweaver and J.K. Webb, {\em Solutions to the tethered galaxy problem in an expanding universe and the observation of receding blueshifted objects,
Am. J. Phys.} {\bf 71} (2003) 358  [astro-ph/0104349].
\bibitem{dec2}
A.B. Whiting, {\em The expansion of space: free-particle motion and the cosmological redshift, The Observatory} {\bf 124} (2004) 174  [astro-ph/0404095].
\bibitem{dec3}
L.A. Barnes, M.J. Francis, J.B. James and G.F. Lewis, {\em Joining the Hubble Flow: implications for expanding space, MNRAS} {\bf 373} (2006) 382 [astro-ph/0609271].
\bibitem{new3}
R.P. van der Marel, G. Besla, T.J. Cox, S.T. Sohn and J. Anderson, {\em The M31 Velocity Vector. III. Future Milky Way-M31-M33 Orbital Evolution, Merging, and Fate of
the Sun}; [astro-ph/1205.6865].
\bibitem{new1}
S.T. Sohn, J. Anderson and R.P. van der Marel, {\em The M31 Velocity Vector. I. Hubble Space Telescope Proper Motion Measurements}; [astro-ph/1205.6863].
\bibitem{new2}
R.P. van der Marel, M. Fardal, G. Besla, R.L. Beaton, S.T. Sohn, J. Anderson, T. Brown and P. Guhathakurta, {\em The M31 Velocity Vector. II. Radial Orbit Towards the
Milky Way and Implied Local Group Mass}; [astro-ph/1205.6864].


\end{thebibliography}
\end{document}